\documentclass[lettersize,journal]{IEEEtran}

\usepackage{amsmath,amssymb,amsfonts,bm}

\usepackage{graphicx}
\usepackage[table]{xcolor} 

\usepackage{array}
\usepackage{multirow}
\usepackage{makecell}
\usepackage{tabularx}
\usepackage{booktabs}
\usepackage{balance}
\usepackage[linesnumbered,vlined,ruled,noend]{algorithm2e}
\SetKwFor{For}{for}{do}{end}
\SetKwIF{If}{ElseIf}{Else}{if}{then}{else if}{else}{end}

\usepackage[caption=false,font=normalsize,labelfont=sf,textfont=sf]{subfig}

\usepackage{textcomp}
\usepackage{verbatim}
\usepackage{enumitem}
\usepackage{float}
\usepackage{url}
\usepackage{cite}
\usepackage[hidelinks]{hyperref}
\usepackage{caption}
\usepackage{stfloats}

\usepackage{adjustbox}
\usepackage{threeparttable}
\usepackage[table]{xcolor}
\AtBeginDocument{%
  \providecommand\BibTeX{{%
    \normalfont B\kern-0.5em{\scshape i\kern-0.25em b}\kern-0.8em\TeX}}}

\begin{document}

\begin{sloppypar}

\title{Mitigating Filter Bubble from the Perspective of Community Detection: A Universal Framework}

\author{Ming~Tang,
        Xiaowen~Huang$^{\ast}$,
        and~Jitao~Sang
        
    \IEEEcompsocitemizethanks{
        \IEEEcompsocthanksitem Ming Tang, Xiaowen Huang, and Jitao Sang are with the School of Computer Science and Technology, Beijing Jiaotong University; the Beijing Key Laboratory of Traffic Data Mining and Embodied Intelligence; and the Key Laboratory of Big Data \& Artificial Intelligence in Transportation, Ministry of Education.
        (E-mail: \href{mailto:23120413@bjtu.edu.cn}{23120413@bjtu.edu.cn}; \href{mailto:xwhuang@bjtu.edu.cn}{xwhuang@bjtu.edu.cn}; \href{mailto:jtsang@bjtu.edu.cn}{jtsang@bjtu.edu.cn})
        }
         
    \thanks{$^{\ast}$corresponding author}

}

\markboth{Journal of \LaTeX\ Class Files,~Vol.~14, No.~8, August~2021}%
{Shell \MakeLowercase{\textit{et al.}}: A Sample Article Using IEEEtran.cls for IEEE Journals}

\maketitle

\begin{abstract}
In recent years, recommender systems have primarily focused on improving accuracy at the expense of diversity, which exacerbates the well-known filter bubble effect. This paper proposes a universal framework called CD-CGCN to address the filter bubble issue in recommender systems from a community detection perspective. By analyzing user-item interaction histories with a community detection algorithm, we reveal that state-of-the-art recommendations often focus on intra-community items, worsening the filter bubble effect. CD-CGCN, a model-agnostic framework integrates a \textbf{C}onditional \textbf{D}iscriminator and a \textbf{C}ommunity-reweighted \textbf{G}raph \textbf{C}onvolutional \textbf{N}etwork which can be plugged into most recommender models. Using adversarial learning based on community labels, it counteracts the extracted community attributes and incorporates an inference strategy tailored to the user’s specific filter bubble state. Extensive experiments on real-world datasets with multiple base models validate its effectiveness in mitigating filter bubbles while preserving recommendation quality. Additionally, by applying community debiasing to the original test set to construct an unbiased test set, we observe that CD-CGCN demonstrates superior performance in capturing users' inter-community preferences.

\end{abstract}


\section{INTRODUCTION}
\label{sec:introduction}
\IEEEPARstart {R}{ecommender} systems have become essential in addressing the issue of information overload \cite{01info} across various domains like social networks \cite{social_network}, online videos \cite{giu2016online_video}, online news \cite{wu2019online_news} and e-commerce \cite{wang18e-commerce}. Recommendation algorithms predict user preferences based on interaction history to provide personalized recommendations \cite{17NCF}. However, almost all of the recommender systems prioritize high accuracy as a means to enhance user satisfaction, which comes at the expense of recommendation diversity, leading to the notorious filter bubble effect \cite{14FB, 21FB, McKay}.

\textit{Filter Bubble} was originally proposed in the context of search engines \cite{Pa11FB}: where two users searching for the same keyword might receive completely different results, implying that each user exists within a "filter bubble"—a personalized information universe. In the field of social sciences, this phenomenon is similar to group multipolarization, where physical isolation in the past prevented information from crossing between distinct groups \cite{23netpo}. Today, this effect also occurs on social media, where platforms often recommend popular content within users' existing social networks \cite{media}, contributing to social group polarization and negative societal impacts \cite{drake16fb}. In transition to recommender systems, current researchers typically attribute the filter bubble effect solely to a reduction in the diversity of recommendation lists \cite{14FB, under_EC}. Their approaches focus on measuring the diversity within each user's recommendation list and considering the average diversity in isolation \cite{gao22miFB, wang22FB, 22mifb}. However, we argue that the filter bubble effect is more of a holistic phenomenon, requiring a closer connection between users and items for a more comprehensive assessment. This phenomenon is particularly evident in communities, as members often share similar backgrounds, interests, and perspectives, making it easier to form a homogenization of information. Based on this, we define and mitigate the filter bubble effect from a community perspective accordingly.
        \begin{figure}[t]
          \centering
          \includegraphics[width=1\linewidth]{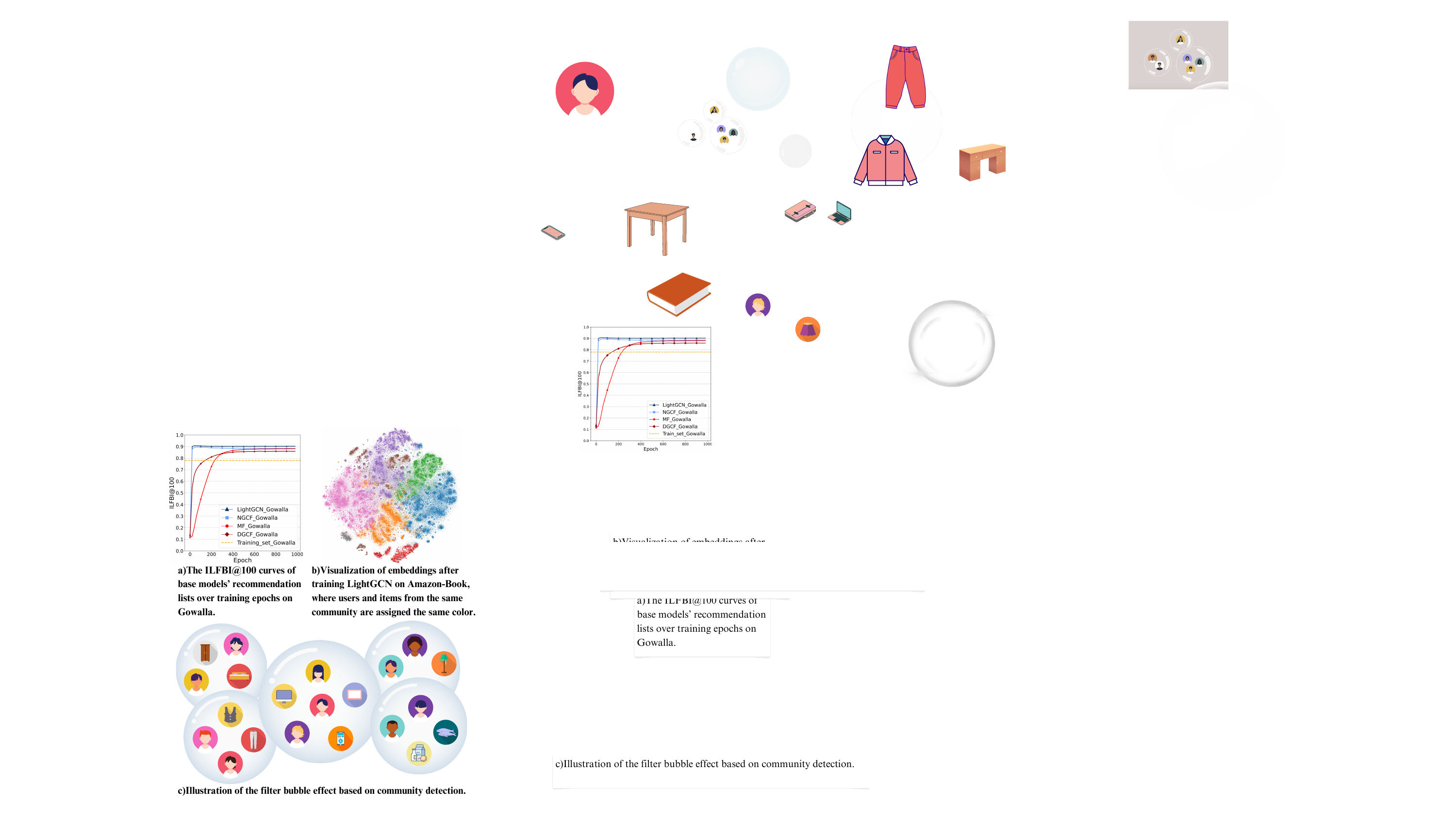}
          \caption{Illustration of the filter bubble effect based on community detection.}
          \label{figure:bubble}
        \end{figure}
In our work, we first apply the widely used community detection algorithm Louvain \cite{08louvain} to obtain the community labels for users and items from the user-item bipartite interaction graph. Next, user's filter bubble index is defined as the proportion of items in the recommendation list that belong to the user’s own community (i.e., intra-community items), and the higher average filter bubble index across all users indicates a more pronounced filter bubble effect (see definitions in Sec. \ref{section:FB_metrics}). Then our findings reveal that current state-of-the-art recommender systems tend to recommend intra-community items to users, and both user and item embeddings exhibit a strong clustering phenomenon within the same community (see details in Sec. \ref{section:motivation}), and due to the existence of the feedback loop in the recommender system \cite{19loop, 20loop}, the filter bubble effect will intensify after multiple rounds of recommendations, leading to the phenomenon of multipolarization among communities \cite{po_so}. As shown in Fig. \ref{figure:bubble}, users and items are isolated within individual bubbles, making it difficult for users to access items outside of their own bubble.

Based on the phenomenon we have observed, in order to mitigate the filter bubble effect which can be exacerbated by the recommendation feedback loop, we propose a universal framework called CD-CGCN with a \textbf{C}onditional \textbf{D}iscriminator and a \textbf{C}ommunity-reweighted \textbf{G}raph \textbf{C}onvolutional \textbf{N}etwork. CD-CGCN can be applied to a wide range of existing recommendations and improve their performance which utilizes adversarial learning to combat the clustering of intra-community embeddings, thereby mitigating the filter bubble effect. Specifically, we feed the user and item embeddings generated by the base model into the Community-reweighted Graph Convolutional Network to extract community attributes and use community labels to train the Conditional Discriminator, which adversarizes the community embeddings of users and items to mitigate their community attributes. Then we introduce Community-enhanced Negative Sampling to further push apart embeddings within the same community. By incorporating a Gradient Reversal Layer, we simplify the alternating training task into an end-to-end training task. During the inference stage, we adaptively fuse the CD-CGCN with the base model based on the user's specific interaction history filter bubble state. 

To summarize, the main contributions are as follows:
        \begin{itemize}[leftmargin=8pt]
        \item We explore the filter bubble effect in recommender systems from a novel community-based perspective and reveal that SOTA recommendations over-recommend items from user's own community. We support this analysis through training curves and embeddings visualization, and we also track the extent of the filter bubble effects experienced by users with different behavior. (Section \ref{section:motivation})
        \item We propose a model-agnostic framework CD-CGCN utilizing Conditional Discriminator and Community-reweighted Convolutional Network to mitigate the filter bubble effect by adversarizing community attributes, which also includes Community-enhanced Negative Sampling, User-adaptive Inference Strategy, and a Gradient Reversal Layer specially introduced to enable end-to-end training. (Section \ref{section:methodology})
        \item We design two novel metrics for evaluating the filter bubble effect based on a community perspective and validate the effectiveness of CD-CGCN across base models on real-world datasets compared with other baselines, and further comparing the performance on the debiased test sets with that on the original test sets demonstrates its unique ability to capture user's inter-community preferences. (Section \ref{section:experiment})
        \end{itemize}

\section{MOTIVATION}
\label{section:motivation}
In this section, we analyze the filter bubble effect in existing recommender systems.

\begin{figure}[t]
          \centering
          \includegraphics[width=1\linewidth]{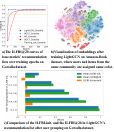}
          \caption{Evidences of the filter bubble effect in recommender systems.}
          \label{figure:visual_before}
        \end{figure}

Community detection aims to identify communities of nodes in a network where nodes within the same community are closely connected to each other, while having fewer connections with nodes outside the community. We put the user-item bipartite interaction graph into the Louvain community detection algorithm \cite{08louvain} to get their community labels, and we find that 78\% of the user's interaction history on the Gowalla \cite{gowalla} dataset are intra-community items. However, 90\% of the items in the top-20 recommendation lists generated by the base models (see details in Sec. \ref{section:base models}) are intra-community items. As shown in Fig. \ref{figure:visual_before}(a), we track their training process and observe the recommendation list's ILFBI@20 (Intra-List Filter Bubble Index@$k$, we define this metric to represent the degree of the filter bubble effect in the top-$k$ recommendation list, and a higher ILFBI indicates a stronger filter bubble effect. For specific definitions, see Sec. \ref{section:FB_metrics}) of LightGCN \cite{lightgcn} and NGCF \cite{NGCF} remains around 0.9 since the beginning of training, while MF \cite{MF} and DGCF \cite{DGCF} show a slower increase but eventually reach similar levels, and their accuracy metrics converge around the 900th epoch, which indicates that the filter bubble effect is related to the recommender model itself, as it tends to consistently recommend a large number intra-community items and then gradually capture the items that the user is interested in.
 
 Afterward, we train LightGCN on Amazon-Book \cite{amazon} and save the embeddings of users and items under the best model, then visualize these embeddings in a two-dimensional space using t-SNE \cite{tsne} which gives the same color to users and items in the same community in Fig. \ref{figure:visual_before}(b). We clearly observe a clustering phenomenon in the embeddings of users and items from the same community, indicated by the same color, which means closely connected users and items within the same community exhibit clustering in their embedding representations after being trained by LightGCN, that is, items are more likely to be recommended to users who belong to the same community. This clustering phenomenon can also be explained by the collaborative filtering principle \cite{01info}. It is important to clarify that our paper does not challenge this fundamental idea, but rather aims to mitigate the filter bubble effect caused by collaborative filtering while still recommending items that align with the user's interests. Thus, we demonstrate that existing recommender systems suffer from the filter bubble effect and the observed community clustering may lead to community multipolarization \cite{po_so} due to the feedback loop in recommender systems \cite{20loop}.
 
Additionally, we categorize users into five groups based on their ILFBI-init. (the proportion of intra-community items in the user's training set) to track the differences between their recommendation lists and training sets on Gowalla dataset. As shown in Fig. \ref{figure:visual_before}(c), we calculate the average ILFBI-init. from the training set and the average ILFBI@20 from the top-20 recommendation list generated by the trained LightGCN for each group, then we calculate their difference which represents the increment of ILFBI.
Our analysis reveals that the increment of ILFBI increases with the ILFBI-init. (since ILFBI has a maximum value of 1, the increment in group (0,8,1.0] is not significant), which means that if a user prefers interacting with inter-community items (with low ILFBI-init.), the recommendation list will include more inter-community items than the training set (the explanation of the increment -0.0492). Conversely, the more a user prefers interacting with intra-community items (with higher ILFBI-init.), the larger the increment in the proportion of intra-community items within their recommendation list relative to the training set. In other words, a higher ILFBI-init. indicates a more pronounced filter bubble effect in the output of the recommender system.

Therefore, we aim to alleviate the clustering phenomenon of user and item embeddings within the same community to encourage recommender systems to recommend more inter-community items, which will reduce the ILFBI of the recommendation list and mitigate the filter bubble effect.

        \begin{figure*}[h]
          \centering
          \includegraphics[width=1\linewidth]{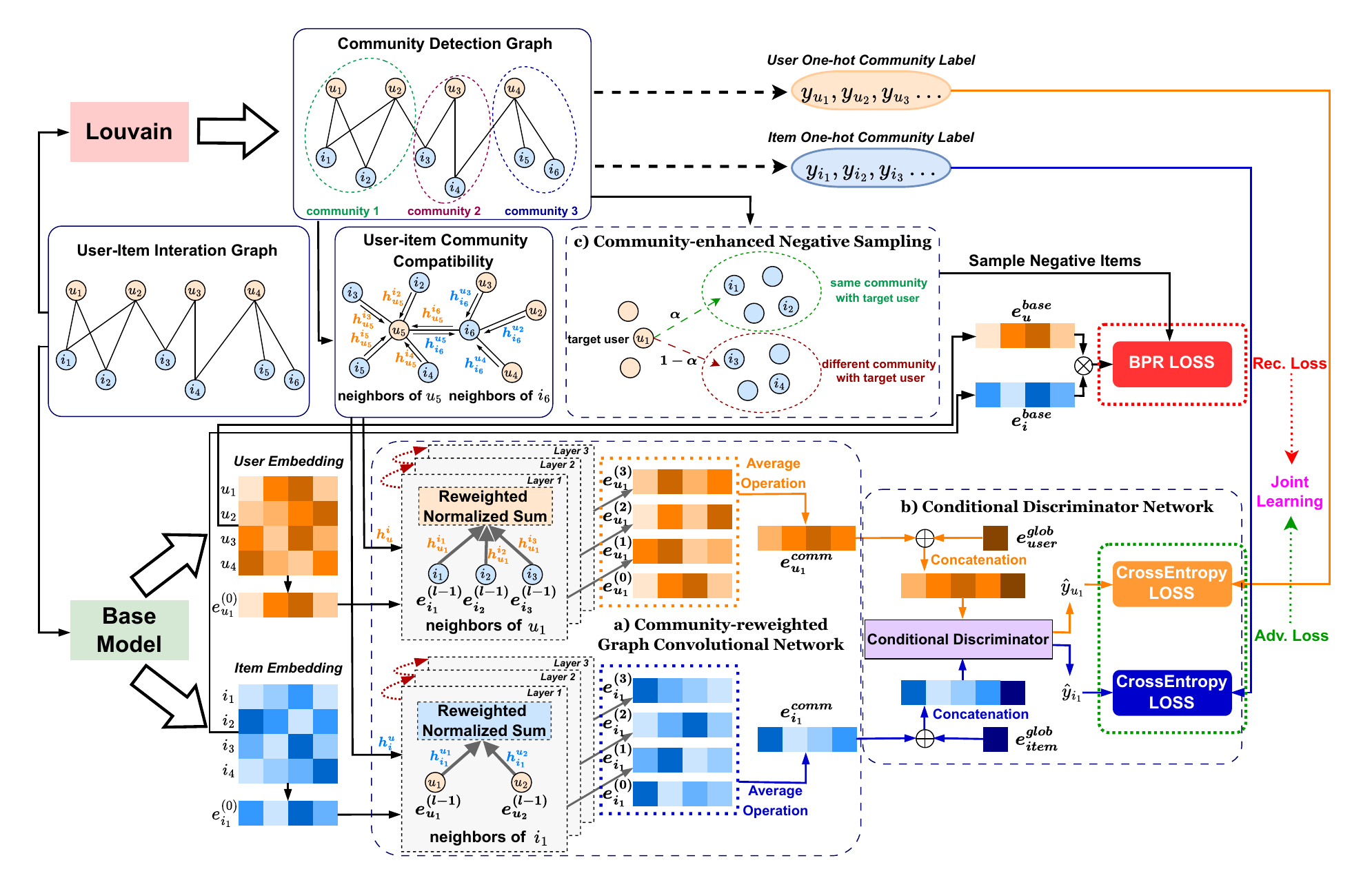}
          \caption{The overall architecture of the CD-CGCN with three core components, which utilizes adversarial learning to mitigate the filter bubble effect.}
          \label{figure:CD-CGCN}
        \end{figure*}
\section{METHODOLOGY}
\label{section:methodology}
    
    In this section, we outline the overall CD-CGCN framework (see Fig. \ref{figure:CD-CGCN}) and a user-adaptive inference strategy.
    
    \subsection{Overview}
    \label{section:overview}
    Formally, we suppose that $\mathcal{U}=\{u_1, u_2, \dots, u_m\} $ $(|\mathcal{U}|=m)$ and $\mathcal{I}=\{i_1, i_2, \dots, i_n\}$ $(|\mathcal{I}|=n)$ denote the user set and the item set, respectively, where $|\cdot|$ is the number of elements in the set. The interaction matrix between the user set and the item set is represented as $\bm{R}\in \mathbb{R}^{m \times n}$, where $R_{u,i}=1$ if the user $u$ has interacted with item $i$ and $R_{u,i}=0$ otherwise. Then we put the user-item bipartite interaction graph $\mathcal{G}=(\mathcal{U} \cup \mathcal{I}, \bm{R})$ into the Louvain community detection algorithm \cite{08louvain} to derive the community label set $\mathcal{C}=\{C_{u_1}, \dots, C_{u_m}, C_{i_1}, \dots, C_{i_n}\} $ $(|\mathcal{C}|=m+n)$ for all users and items, where $C_{u_1}=2$ means user $u_1$ belongs to Community-2 and similarly for items. The goal of CD-CGCN is to generate a set of items $\mathcal{T}_u$ for each user $u$ on the basis of each base model, aiming to achieve better accuracy and filter bubble metrics compared with the original base model and other baselines. 
    
    As shown in Fig. \ref{figure:CD-CGCN}, CD-CGCN consists of three core parts. In detail, first, we use base embeddings of users and items from the base model (BM) to train the main recommendation task through BPR Loss \cite{bpr}, meanwhile, we feed base embeddings into the Community-reweighted Graph Convolutional Network (CGCN) to obtain their community embeddings which include community attributes. In order to alleviate the community aggregation phenomenon of embeddings as mentioned in Sec. \ref{section:motivation}, we use the community embeddings to train the Conditional Discriminator (CD) based on community labels of user and item through Cross-Entropy Loss \cite{CELoss}. Finally, we design a Community-enhanced Negative Sampling Strategy and a User-adaptive Inference Strategy to further improve the performance, and utilize a Gradient Reversal Layer to transform the training mode.
    
    \subsection{Community-reweighted Graph Convolutional Network}
    \label{section:CGCN}
    We first introduce the calculation of user-item community compatibility $\bm{h}^i_u$ and $\bm{h}^u_i$. For a user $u$, the community compatibility of neighbor item $i$ for the user is defined as follows:
    {
    \begin{equation}
    h^i_u = \frac{\Big|\{k \mid R_{u,k}=1, C_k=C_i, k \in \mathcal{I}\}\Big|}{\Big|\{k \mid R_{u,k}=1, k \in \mathcal{I}\}\Big|}
    \label{equation:0}
    \end{equation}
    }where $C_k$ (mentioned in Sec. \ref{section:overview}) represents the community to which item $k$ belongs, and this expression indicates the proportion of items in user $u$'s interaction history that belong to the same community as item $i$.

    For example, after assigning community labels to users and items using a community detection algorithm, suppose that there are three communities labeled 1, 2, and 3. If a user $u_1$ totally interacted with eight Community-1 items and two Community-2 items, and if $u_1$'s neighbor item $i_1$ belongs to Community-1, then we set $\bm{h}^{i_1}_{u_1}$ to 0.8. The definition of $\bm{h}^u_i$ is similar. Upon analyzing real-world datasets, we further observe that when user $u$ and his neighbor item $i_1$ belong to the same community, $\bm{h}^{i_1}_u$ tends to be higher than $\bm{h}^{i_2}_u$, where another neighbor item $i_2$ belongs to a different community, and the situation is the same for $\bm{h}^u_i$.
    
    After getting user-item community compatibility $\bm{h}^i_u$ and $\bm{h}^u_i$, we propose Community-reweighted Graph Convolutional Network to extract the community attributes from base embeddings (see Fig. \ref{figure:CD-CGCN}(a)). Let the base embeddings of user $u$ and item $i$ obtained from the base model be represented as $\bm{e}_u^{base}$ and $\bm{e}_i^{base}$, respectively. Then we initialize Layer-0 embeddings as $\bm{e}_u^{(0)}$ = $\bm{e}_u^{base}$ and $\bm{e}_i^{(0)}$ = $\bm{e}_i^{base}$, and the reweighted propagation method using the user-item interaction graph is as follows:
    
    {
    \begin{equation}
    \bm{e}_u^{(k+1)} = \sum_{i \in \mathcal{N}_u} \frac{h^i_u}{\sqrt{\smash[b]{\sum_{i \in \mathcal{N}_u} h^i_u}}  \sqrt{\smash[b]{\sum_{u \in \mathcal{N}_i} h^u_i}}} \bm{e}_i^{(k)}
    \label{equation:1}
    \end{equation}
    }
    
    {
    \begin{equation}
    \bm{e}_i^{(k+1)} = \sum_{u \in \mathcal{N}_i} \frac{h^u_i}{\sqrt{\smash[b]{\sum_{u \in \mathcal{N}_i} h^u_i}}  \sqrt{\smash[b]{\sum_{i \in \mathcal{N}_u} h^i_u}}} \bm{e}_u^{(k)}
    \label{equation:2}
    \end{equation}
    }where $\bm{e}_u^{(k)}$ and $\bm{e}_i^{(k)}$ respectively denote the Layer-$k$ embeddings of user $u$ and item $i$ after $k$ layers propagation, $\mathcal{N}_u$ denotes the set of items interacted with by user $u$, $\mathcal{N}_i$ denotes the set of users who interact with item $i$. The symmetric reweighted normalization term \(\frac{1}{\sqrt{\sum_{u \in \mathcal{N}_i} h^u_i\vphantom{\sum_{i \in \mathcal{N}_u}}} \sqrt{\sum_{i \in \mathcal{N}_u} h^i_u}}\) is transformed from the standard GCN \cite{17GCN}.

    As mentioned before, if user $u$ and item $i$ belong to the same community, $\bm{h}^i_u$ and $\bm{h}^u_i$ will be large and close to 1, whereas they will be small and close to 0 when user $u$ and item $i$ belong to different communities. Therefore, as the layer becomes deeper, the embeddings of users and items within the same community tend to exhibit a strong and remarkable clustering phenomenon, which means CGCN extracts the community attributes from base embeddings to better serve our next stage of adversarial learning. 
    
    By propagating total $L$ layers, CGCN obtains $L + 1$  embeddings to describe the community attributes of a user $u$ : $(\bm{e}^{(0)}_u, \bm{e}^{(1)}_u, \dots, \bm{e}^{(L)}_u)$ and an item $i$ : $(\bm{e}^{(0)}_i, \bm{e}^{(1)}_i, \dots, \bm{e}^{(L)}_i)$. Then we perform an average operation on these $L$ + 1 embeddings obtained at each layer to form the final community embedding which represents the community attributes of the user or item:
    {
    \begin{equation}
    \bm{e}_u^{comm} = \sum^L_{k=0} \bm{e}_u^{(k)}\quad ; \quad\bm{e}_i^{comm} = \sum^L_{k=0} \bm{e}_i^{(k)}
    \label{equation:3}
    \end{equation}
    }
    
    In summary, $\bm{e}_u^{comm}$ and $\bm{e}_i^{comm}$ effectively capture community attributes while preserving base embeddings, and they are obtained through weighted convolution of the CGCN, which operates without any learnable parameters.

    \subsection{Conditional Discriminator Network}
    \label{section:CD}
    In this part, we design a Conditional Discriminator Network denoted as $\bm{\mathcal{D}^{condi}}$ which consists of two fully-connected layers with parameters $\theta_d$ (see Fig. \ref{figure:CD-CGCN}(b)). 
    
    First, to help the discriminator distinguish whether $\bm{e}^{comm}$ originates from user's or item's community embedding, we introduce a global user embedding $\bm{e}_{user}^{glob}$ and a global item embedding $\bm{e}_{item}^{glob}$ shared across all users and items, respectively.

     We concatenate a user $u$'s community embedding $\bm{e}_{u}^{comm}$ and the global user embedding $\bm{e}_{user}^{glob}$ together, and then feed them into the Conditional Discriminator Network $\bm{\mathcal{D}^{condi}}$ to get the user $u$'s community probability distribution representation. The same process applies to item $i$, and the formulas are shown below:
     {
     \begin{equation}
     \hat{y}_{u} = \bm{\mathcal{D}^{condi}}( \bm{e}_{u}^{comm} || \bm{e}_{user}^{glob}; \theta_d)
     \label{equation:4}
     \end{equation}
     }
     {
     \begin{equation}
     \hat{y}_{i} = \bm{\mathcal{D}^{condi}}( \bm{e}_{i}^{comm} || \bm{e}_{item}^{glob}; \theta_d)   
     \label{equation:5}
     \end{equation}
     }
    
    After getting $\hat{y}_{u}$ and $\hat{y}_{i}$, we utilize them to compute the cross-entropy loss using the one-hot community labels $y_u$ and $y_i$ derived from the community labels $C_u$ and $C_i$ (mentioned in Sec. \ref{section:overview}). Given the community probability distribution $\hat{y}$ and the one-hot community label $y$, the Cross-Entropy Loss is defined as follows:
    \begin{equation}
    CE(\hat{y},y) = -\sum_{k=1}^n y_klog\hat{y}_k    
    \label{equation:6}
    \end{equation}
    where $n$ represents the number of dimensions of $y$ and $\hat{y}$, and here it denotes the total number of communities.

    \subsection{Community-enhanced Negative Sampling}
    \label{section:neg_sample}

    In this part, we propose a negative sampling strategy to better serve our objective. Traditional recommender systems randomly select items from a user's non-interacted items as negative samples. In our CD-CGCN framework, after being given a user $u$ with positive sample $i$, we aim to prioritize selecting non-interacted items that belong to $u$'s community as a negative sample $j$, further increasing the distance of representation within the same community. Our Community-enhanced Negative Sampling strategy is shown in Fig. \ref{figure:CD-CGCN}(c): we sample non-interacted items from the user's same community as the negative sample with a probability of $\alpha$, while with a probability of 1-$\alpha$ we directly sample from the pool of non-interacted items, where $\alpha$ is a hyper-parameter to control the intensity of community-enhanced negative sampling. This strategy increases the likelihood of recommending inter-community items by enhancing the probability of sampling intra-community items as negative samples, which effectively mitigates the filter bubble effect.

    \subsection{Model Optimization}
    Now we focus on the input training triplet set $\mathcal{S}$ $\subseteq$ \{$(u, i, j)|u\in\mathcal{U}$,$\:i,j\in\mathcal{I}, R_{u,i}=1, R_{u,j}=0$\}, that is, for an interaction record ($u$,$i$), we sample an item $j$ as a negative item to form a triplet ($u$, $i$, $j$), and the negative sampling strategy is mentioned in Sec. \ref{section:neg_sample}.

    \subsubsection{Recommendation Loss.}
    \label{section:3.5.1}
    We utilize $\bm{e}_u^{base}$ and $\bm{e}_i^{base}$ obtained from base models to compute Bayesian Personalized Ranking (BPR) \cite{bpr} as the recommendation loss:
    {
    \begin{equation}
    \mathcal{L}_{rec}(u,i,j;\theta_b) = -\sum_{(u,i,j)\in\mathcal{S}}ln\:\sigma(s(\bm{e}_u^{b}, \bm{e}_i^{b})-s(\bm{e}_u^{b}, \bm{e}_j^{b}))
    \label{equation:7}
    \end{equation}
    }where $s$($\bm{e}_u$,$\bm{e}_i$) = $\bm{e}_u^T$$\bm{e}_i$ represents the user-item score predicted via inner product, we substitute $\bm{e}_u^{b}$/$\bm{e}_i^{b}$ for $\bm{e}_u^{base}$/$\bm{e}_i^{base}$, and $\theta_b$ represents the parameters in base model which can be regarded as $\bm{e}_u^{base}$ and $\bm{e}_i^{base}$. In addition, CD-CGCN can also be adapted to other recommendation tasks by replacing the corresponding main recommendation loss.
    
    \subsubsection{Adversarial Loss.}
    We use the community probability distribution $\hat{y}$ output from the Conditional Discriminator Network and the one-hot community label $y$ to compute the adversarial loss:
    {
    \begin{equation}
    \begin{split}
    \mathcal{L}_{adv}(u,i,j;\theta_b,\theta_d) = \\ \sum_{(u,i,j) \in \mathcal{S}}  CE(\hat{y}_u, y_u)& + CE(\hat{y}_i, y_i) + CE(\hat{y}_j, y_j)
    \label{equation:8}
    \end{split}
    \end{equation}
    }where $CE(\hat{y}, y)$ is the Cross-Entropy Loss defined in Equation \ref{equation:6}, $\hat{y}_u$ and $\hat{y}_i$ come from Equations \ref{equation:4} and \ref{equation:5}, and $\theta_d$ represents the parameters in the Conditional Discriminator Network (including global user/item embedding $\bm{e}_{user}^{glob}$ and $\bm{e}_{item}^{glob}$).
    \subsubsection{Parameter Updating.}
    \label{section:beta}
    We begin by using $\mathcal{L}_{rec}(\theta_b)$ and $\mathcal{L}_{adv}(\theta_b,\theta_d)$ as simplified representations for $\mathcal{L}_{rec}(u,i,j;\theta_b)$ and $\mathcal{L}_{adv}(u,i,j;\theta_b,\theta_d)$. Next, we combine the above two losses to define our total loss $\mathcal{L}(\theta_b,\theta_d)$:
    \begin{equation}
    \mathcal{L}(\theta_b,\theta_d) = \mathcal{L}_{rec}(\theta_b) - \beta \mathcal{L}_{adv}(\theta_b,\theta_d)
    \label{equation:9}
    \end{equation}where $\beta$ is a hyper-parameter to control the intensity of adversarial learning.
    
    We update $\theta_b$ and $\theta_d$ alternately as follows:
    \begin{equation}
    {\theta}_b^* = \arg\min_{\theta_b} \mathcal{L}(\theta_b,\hat{\theta}_d)
    \label{equation:10}
    \end{equation}
    \begin{equation}
    {\theta}_d^* = \arg\max_{\theta_d} \mathcal{L}(\hat{\theta_b},\theta_d)
    \label{equation:11}
    \end{equation}
    Equation \ref{equation:10} updates $\theta_b$ to minimize $\mathcal{L}$ while keeping $\theta_d$ fixed as $\hat{\theta}_d$, with the aim of increasing the gap between the user's scores for positive and negative samples by lowering $\mathcal{L}_{rec}(\theta_b)$ and making it more difficult for the Conditional Discriminator to determine the user/item's community label by raising $\mathcal{L}_{adv}(\theta_b,\hat{\theta}_d)$.

    Equation \ref{equation:11} updates $\theta_d$ to maximize $\mathcal{L}$ while keeping $\theta_b$ fixed as $\hat{\theta}_b$, with the aim of updating the Conditional Discriminator's parameters to better determine the user/item's community label by lowering $\mathcal{L}_{adv}(\hat{\theta_b},\theta_d)$.

    Through the alternating training process above: 1) $s(\bm{e}_u^{b}, \bm{e}_i^{b})$ better captures the user $u$'s preference for the item $i$. 2) $\bm{e}_{u}^{comm}$ and $\bm{e}_{i}^{comm}$ with community attributes extracted from $\bm{e}_u^{b}$ and $\bm{e}_i^{b}$ make it more difficult for the Conditional Discriminator to identify their community labels. 3) The discrimination ability of the Conditional Discriminator is getting stronger and stronger. In short, in addition to learning better representation of $\bm{e}_u^{b}$ and $\bm{e}_i^{b}$, $\bm{e}_{u}^{comm}$ and $\bm{e}_{i}^{comm}$ with community attributes are no longer as easily distinguishable as they used to be.
    
    \subsubsection{Training Form Simplified.}
    As mentioned above, Equations \ref{equation:9}, \ref{equation:10} and \ref{equation:11} describe the entire alternating training process.
    Considering that this training form is complex and may lead to convergence instability \cite{liu2020loss}, we can simplify it by introducing a special gradient reversal layer (GRL) \cite{GRL} $\mathcal{R}_\beta$ with a single hyper-parameter $\beta$ (mentioned in Equation \ref{equation:9}) defined as follows:
    \begin{equation}
    \mathcal{R}_\beta(\bm{x}) = \bm{x} \:\:\: ;\:\:\:  \frac{d\mathcal{R}_\beta(\bm{x})}{d\bm{x}} = -\beta\bm{I}
    \label{equation:12}
    \end{equation}

    With this GRL $\mathcal{R}_\beta$, we transform the $\mathcal{L}(\theta_b,\theta_d)$ in Equation \ref{equation:9} into $\mathcal{L}^{\mathcal{R}_\beta}(\theta_b,\theta_d)$:
    \begin{equation}
    \mathcal{L}^{\mathcal{R}_\beta}(\theta_b,\theta_d) = \mathcal{L}_{rec}(\theta_b) + \mathcal{L}_{adv}(\mathcal{R}_\beta(\theta_b),\theta_d)
    \label{equation:14}
    \end{equation}

    Then we can give the new parameter update method:
    \begin{equation}
    {\theta}_b^*, {\theta}_d^* = \arg\min_{\theta_b, \theta_d} \mathcal{L}^{\mathcal{R}_\beta}(\theta_b,\theta_d)
    \label{equation:15}
    \end{equation}

    With GRL $\mathcal{R}_\beta$ defined as Equation \ref{equation:12}, Equations \ref{equation:14} and \ref{equation:15} are equivalent to Equations \ref{equation:9}, \ref{equation:10} and \ref{equation:11} (proved in Appendix \ref{App}). In other words, we simplify an unstable alternating training task into a stable end-to-end training task.

    \subsection{User-adaptive Inference Strategy}
    \label{section:user_ada}
    To address the aforementioned issue of user differentiation in Sec. \ref{section:motivation}, we design an inference strategy to adaptively fuse the prediction scores from the base model and the CD-CGCN. We define the weight for user $u$ as follows:
    \begin{equation}
    \eta_u = \frac{ILFBI_u^{init.}}{2*\overline{ILFBI_u^{init.}}}
    \label{equation:16}
    \end{equation} where $ILFBI_u^{init.}$ represents user $u$'s ILFBI-init. in the training set, {\footnotesize $\overline{ILFBI_u^{init.}}$} means the average of $ILFBI_u^{init.}$ for all users.

    In this formula, $\eta_u$ is directly proportional to $ILFBI_u^{init.}$, with normalization achieved through the denominator. Revisiting our strategy, we want CD-CGCN to contribute more to users with higher $ILFBI_u^{init.}$, formally,
    \begin{equation}
    Y_{u,i} = \eta_u * Y_{u,i}^{CD} + (1-\eta_u) * Y_{u,i}^{BM}
    \label{equation:fuse}
    \end{equation}where $Y_{u,i}$ is the inference score for user $u$ and item $i$, $Y_{u,i}^{CD}=s(\bm{e}_u^{base}, \bm{e}_i^{base})$ represents the prediction score from CD-CGCN mentioned above, and $Y_{u,i}^{BM}$ refers to the prediction score from the pretrained base model.

    From Equation \ref{equation:fuse}, we can see that inference for users with large $ILFBI_u^{init.}$ will rely more on $Y_{u,i}^{CD}$, which alleviates the filter bubble effect for these users to a greater extent.

\section{EXPERIMENTS}
\label{section:experiment}
    In this section, we conduct extensive experiments to evaluate our CD-CGCN framework and answer the following research questions:
    \begin{itemize}[leftmargin=8pt]
     \item \textbf{RQ1:} How does CD-CGCN perform on classical and state-of-the-art base models compared with other baselines?  
     \item \textbf{RQ2:} How does CD-CGCN mitigate the filter bubble effect throughout the training process? And what changes are observed in the visualization of embeddings? (compared with Sec. \ref{section:motivation})  
     \item \textbf{RQ3:} How do different components of CD-CGCN contribute to its performance?
     \item \textbf{RQ4:} We further find community bias in the test sets of real-world datasets. Does CD-CGCN perform better on the debiased test sets?
     \item \textbf{RQ5:} How do the key hyper-parameters (i.e., $\alpha$ in Sec. \ref{section:neg_sample} and $\beta$ in Sec. \ref{section:beta}) influence the performance of CD-CGCN?
    \end{itemize}

    \subsection{Experimental Settings}
        \begin{table}[!h]
          \caption{Statistics of datasets.}
          \label{table:dataset}
          \centering
          \scalebox{0.9}{
          \begin{tabular}{c|cccccc}
            \toprule
            Dataset &  Users  &  Items  &  Interaction  &  Density &  N. of Com.  \\
            \midrule
            Gowalla &  29,858  &  40,981  & 1,027,370  &  0.00084  &  18 \\
            Amazon-Book & 52,643 & 91,599 & 2,984,108 & 0.00062 & 9\\
            ML-2k &  2,095  &  5,254  & 367,761  &  0.03341  &  6 \\
            \bottomrule
          \end{tabular}
          }
          
        \end{table}
        \subsubsection{Datasets.}

         We utilize three real-world datasets under different recommendation scenarios in our experiments to evaluate the effectiveness of CD-CGCN: Gowalla \cite{gowalla}, Amazon-Book \cite{amazon} and MovieLens-2k \cite{Cantador:RecSys2011}. Table \ref{table:dataset} summarizes the detailed statistics of datasets, and the last column shows the Louvain community detection algorithm \cite{08louvain} divides datasets Gowalla, Amazon-Book and ML-2k into 18, 9 and 6 communities, respectively.
         
        
        \subsubsection{Base Models.}
        \label{section:base models}
         We employ four recommender systems as base models: MF \cite{MF}, NGCF \cite{NGCF},  LightGCN \cite{lightgcn}, and DGCF \cite{DGCF}, which are described as follows:
        \begin{itemize}[leftmargin=20pt]
        \item \textbf{MF \cite{MF}:} A classical collaborative filtering method that uses the inner product of embeddings to capture user's preferences for item.
        \item \textbf{NGCF \cite{NGCF}:} A GCN-based model that exploits high-order connection relationships that aggregates information across the user-item interaction graph.
        \item \textbf{LightGCN \cite{lightgcn}:} A simplified GCN model that simplifies the message-passing rule by linearly aggregating user and item embeddings on the interaction graph to exploit high-order information.
        \item \textbf{DGCF \cite{DGCF}:} A disentangled representation model that decomposes user preferences into multiple latent factors, enhancing the representation capacity of collaborative filtering by modeling fine-grained user intentions.
        \end{itemize}

        \subsubsection{Baselines.}
        \label{section:baselines}
        
        To validate the effectiveness of CD-CGCN, we compare it with three representative baselines under the community context, each focusing on a different aspect: diversity, fairness, and bias mitigation.
        
        \begin{itemize}[leftmargin=20pt]
        
        \item \textbf{MMR \cite{98Carbonell}}:
        We adopt the Maximum Marginal Relevance (MMR) framework to enhance diversity in recommendation lists by encouraging community balance. It reorders the top-ranked items from base models through a greedy strategy that penalizes items from overrepresented communities.
        
        \item \textbf{Fairness \cite{19Beutel}}:
        This baseline introduces a fairness-aware pairwise ranking approach by incorporating a regularization term into the loss function. The term encourages the model to increase embedding distances between items from the same community, thus promoting community-level fairness in recommendations.
        
        \item \textbf{IPS \cite{joachims2017unbiased, schnabel2016recommendations}}:
        Inverse Propensity Scoring (IPS) is applied to reweight the training loss, giving more importance to interactions across different communities. This aims to reduce the bias caused by the prevalence of same-community interactions during training and to mitigate embedding clustering effects.
        
        \end{itemize}
\begin{table*}[!ht]
  \caption{Performance comparison on three real world datasets, the best results are bold-faced and highlighted.}
  \centering
  
  \renewcommand{\arraystretch}{1.05}
  \setlength{\tabcolsep}{5pt}
  \fontsize{9.5}{10.5}\selectfont

  \resizebox{\textwidth}{!}{
    
    \begin{tabular}{|>{\centering\arraybackslash}p{2cm}|p{2.5cm}|>{\centering\arraybackslash}p{1.4cm}>{\centering\arraybackslash}p{1.4cm}>{\centering\arraybackslash}p{1.4cm}>{\centering\arraybackslash}p{1.4cm}>{\centering\arraybackslash}p{1.4cm}| 
               >{\centering\arraybackslash}p{1.4cm}>{\centering\arraybackslash}p{1.4cm}>{\centering\arraybackslash}p{1.4cm}>{\centering\arraybackslash}p{1.4cm}>{\centering\arraybackslash}p{1.4cm}| 
               >{\centering\arraybackslash}p{1.4cm}>{\centering\arraybackslash}p{1.4cm}>{\centering\arraybackslash}p{1.4cm}>{\centering\arraybackslash}p{1.4cm}>{\centering\arraybackslash}p{1.4cm}|}
      \toprule
      \multirow{2}{*}{\parbox{2cm}{\centering \textbf{Models}}} & 
      \multirow{2}{*}{\textbf{Baselines}} & 
      \multicolumn{5}{c|}{\textbf{Gowalla}} & 
      \multicolumn{5}{c|}{\textbf{Amazon-Book}} & 
      \multicolumn{5}{c|}{\textbf{ML-2k}}  \\

        & & \scalebox{0.85}{P@20 $\uparrow$} & \scalebox{0.85}{R@20 $\uparrow$} & \scalebox{0.85}{N@20 $\uparrow$} & \scalebox{0.85}{F@20 $\downarrow$} & \scalebox{0.85}{G@20 $\downarrow$} & 
        \scalebox{0.85}{P@20 $\uparrow$} & \scalebox{0.85}{R@20 $\uparrow$} & \scalebox{0.85}{N@20 $\uparrow$} & \scalebox{0.85}{F@20 $\downarrow$} & \scalebox{0.85}{G@20 $\downarrow$} & 
        \scalebox{0.85}{P@20 $\uparrow$} & \scalebox{0.85}{R@20 $\uparrow$} & \scalebox{0.85}{N@20 $\uparrow$} & \scalebox{0.85}{F@20 $\downarrow$} & \scalebox{0.85}{G@20 $\downarrow$} \\

      \midrule
      \multirow{5}{*}{\parbox{2cm}{\centering MF}} 
      & BPR   & \cellcolor[HTML]{fed9ae}{0.0403} & 0.1361 & 0.1064 & 0.9043 & 0.9322 & 0.0130 & 0.0315 & 0.0239 & 0.7757 & 0.8267 & \cellcolor[HTML]{fed9ae}{0.1864} & \cellcolor[HTML]{fed9ae}{0.1472} & \cellcolor[HTML]{fed9ae}{0.2267} & 0.5919 & 0.6807  \\
      & +MMR   & 0.0400 & \cellcolor[HTML]{fed9ae}{0.1383} & \cellcolor[HTML]{fed9ae}{0.1065} & \cellcolor[HTML]{fed9ae}{0.8827} & \textbf{\cellcolor[HTML]{fbaf5c}{0.9280}} &  \cellcolor[HTML]{fed9ae}{0.0130} & \cellcolor[HTML]{fed9ae}{0.0316} & \cellcolor[HTML]{fed9ae}{0.0239} & 0.7462 & 0.8139 & 0.1818 & 0.1387 & 0.2150 & 0.5415 & \textbf{\cellcolor[HTML]{fbaf5c}{0.6521}}  \\
      & +Fairness  & 0.0394 & 0.1317 & 0.1009 & 0.9201 & 0.9348 & 0.0115 & 0.0275 & 0.0210 & 0.7694 & 0.8251 & 0.1646 & 0.1329 & 0.1917 & 0.5375 & 0.6738  \\
      
      & +IPS  & 0.0393 & 0.1336 & 0.1030 & 0.8890 & 0.9300 & 0.0114 & 0.0276 & 0.0212 & \cellcolor[HTML]{fed9ae}{0.7257} & \cellcolor[HTML]{fed9ae}{0.8086} & 0.1815 & 0.1421 & 0.2153 & \cellcolor[HTML]{fed9ae}{0.5254} & \cellcolor[HTML]{fed9ae}{0.6590}  \\
      
      & +CD-CGCN & \textbf{\cellcolor[HTML]{fbaf5c}{0.0418}} & \textbf{\cellcolor[HTML]{fbaf5c}{0.1402}} & \textbf{\cellcolor[HTML]{fbaf5c}{0.1106}} & \textbf{\cellcolor[HTML]{fbaf5c}{0.8777}} & \cellcolor[HTML]{fed9ae}{0.9293} & \textbf{\cellcolor[HTML]{fbaf5c}{0.0147}} & \textbf{\cellcolor[HTML]{fbaf5c}{0.0362}} & \textbf{\cellcolor[HTML]{fbaf5c}{0.0276}} & \textbf{\cellcolor[HTML]{fbaf5c}{0.6864}} & \textbf{\cellcolor[HTML]{fbaf5c}{0.7990}} & \textbf{\cellcolor[HTML]{fbaf5c}{0.2080}} & \textbf{\cellcolor[HTML]{fbaf5c}{0.1598}} & \textbf{\cellcolor[HTML]{fbaf5c}{0.2588}} & \textbf{\cellcolor[HTML]{fbaf5c}{0.4974}} & 0.6768  \\
      \midrule
      \multirow{5}{*}{\parbox{2cm}{\centering NGCF}}
      & BPR   & 0.0442 & 0.1474 & 0.1172 & 0.9071 & 0.9326 & 0.0140 & 0.0342 & 0.0260 & 0.7872 & 0.8308 & 0.2453 & 0.1879 & 0.3078 & 0.6642 & 0.7411  \\
      & +MMR   & \cellcolor[HTML]{fed9ae}{0.0476} & \textbf{\cellcolor[HTML]{fbaf5c}{0.1622}} & \cellcolor[HTML]{fed9ae}{0.1287} & \cellcolor[HTML]{fed9ae}{0.8791} & \cellcolor[HTML]{fed9ae}{0.9273} &  0.0138 & 0.0341 & 0.0257 & \cellcolor[HTML]{fed9ae}{0.7453} & \textbf{\cellcolor[HTML]{fbaf5c}{0.8139}} & \cellcolor[HTML]{fed9ae}{0.2477} & \cellcolor[HTML]{fed9ae}{0.1900} & \cellcolor[HTML]{fed9ae}{0.3108} & 0.6266 & \cellcolor[HTML]{fed9ae}{0.7187}  \\
      & +Fairness   & 0.0444 & 0.1460 & 0.1167 & 0.9212 & 0.9351 & \cellcolor[HTML]{fed9ae}{0.0145}  & \cellcolor[HTML]{fed9ae}{0.0353} & \cellcolor[HTML]{fed9ae}{0.0268} & 0.7847 & 0.8306 & 0.2431 & 0.1871 & 0.2974 & 0.6182 & \textbf{\cellcolor[HTML]{fbaf5c}{0.7053}}  \\

      & +IPS  & 0.0442 & 0.1477 & 0.1169 & 0.8881 & 0.9303 & 0.0135 & 0.0330 & 0.0251 & 0.7457 & \cellcolor[HTML]{fed9ae}{0.8169} & 0.2374 & 0.1787 & 0.2976 & \cellcolor[HTML]{fed9ae}{0.5925} & 0.7345  \\
      
      & +CD-CGCN & \textbf{\cellcolor[HTML]{fbaf5c}{0.0490}} & \cellcolor[HTML]{fed9ae}{0.1609} & \textbf{\cellcolor[HTML]{fbaf5c}{0.1292}} & \textbf{\cellcolor[HTML]{fbaf5c}{0.8612}} & \textbf{\cellcolor[HTML]{fbaf5c}{0.9264}} & \textbf{\cellcolor[HTML]{fbaf5c}{0.0160}} & \textbf{\cellcolor[HTML]{fbaf5c}{0.0395}} & \textbf{\cellcolor[HTML]{fbaf5c}{0.0301}} & \textbf{\cellcolor[HTML]{fbaf5c}{0.7262}} & 0.8242 & \textbf{\cellcolor[HTML]{fbaf5c}{0.2585}} & \textbf{\cellcolor[HTML]{fbaf5c}{0.1995}} & \textbf{\cellcolor[HTML]{fbaf5c}{0.3277}} & \textbf{\cellcolor[HTML]{fbaf5c}{0.5688}} & 0.7521  \\
      \midrule
      \multirow{5}{*}{LightGCN}
      & BPR   & \cellcolor[HTML]{fed9ae}{0.0559} & 0.1825 & 0.1545 & 0.9091 & 0.9330 & \cellcolor[HTML]{fed9ae}{0.0172} & \cellcolor[HTML]{fed9ae}{0.0416} & \cellcolor[HTML]{fed9ae}{0.0323} & 0.8082 & 0.8402 & \cellcolor[HTML]{fed9ae}{0.2569} & \cellcolor[HTML]{fed9ae}{0.1991} & \cellcolor[HTML]{fed9ae}{0.3292} & 0.6505 & 0.7364  \\
      & +MMR   & 0.0541 & \cellcolor[HTML]{fed9ae}{0.1850} & \cellcolor[HTML]{fed9ae}{0.1550} & 0.8942 & 0.9311 &  0.0170 & 0.0414 & 0.0319 & \cellcolor[HTML]{fed9ae}{0.7438} & \textbf{\cellcolor[HTML]{fbaf5c}{0.8140}} & 0.2555 & 0.1972 & 0.3263 & 0.6194 & \cellcolor[HTML]{fed9ae}{0.7256}  \\
      & +Fairness   & 0.0535 & 0.1735 & 0.1462 & 0.8948 & \cellcolor[HTML]{fed9ae}{0.9310} & 0.0149 & 0.0363 & 0.0277 & 0.7823 & 0.8389 & 0.2451 & 0.1945 & 0.3140 & 0.6421 & \textbf{\cellcolor[HTML]{fbaf5c}{0.7222}}  \\

      & +IPS  & 0.0555 & 0.1813 & 0.1539 & \cellcolor[HTML]{fed9ae}{0.8895} & 0.9311 & 0.0168 & 0.0406 & 0.0315 & 0.7654 & 0.8280 & 0.2563 & 0.1972 & 0.3273 & \cellcolor[HTML]{fed9ae}{0.6168} & 0.7304  \\
      
      & +CD-CGCN & \textbf{\cellcolor[HTML]{fbaf5c}{0.0571}} & \textbf{\cellcolor[HTML]{fbaf5c}{0.1854}} & \textbf{\cellcolor[HTML]{fbaf5c}{0.1577}} & \textbf{\cellcolor[HTML]{fbaf5c}{0.8858}} & \textbf{\cellcolor[HTML]{fbaf5c}{0.9303}} & \textbf{\cellcolor[HTML]{fbaf5c}{0.0184}} & \textbf{\cellcolor[HTML]{fbaf5c}{0.0447}} & \textbf{\cellcolor[HTML]{fbaf5c}{0.0346}} & \textbf{\cellcolor[HTML]{fbaf5c}{0.7398}} & \cellcolor[HTML]{fed9ae}{0.8227} & \textbf{\cellcolor[HTML]{fbaf5c}{0.2626}} & \textbf{\cellcolor[HTML]{fbaf5c}{0.2031}} & \textbf{\cellcolor[HTML]{fbaf5c}{0.3360}} & \textbf{\cellcolor[HTML]{fbaf5c}{0.5943}} & 0.7409  \\
      \midrule
      \multirow{5}{*}{\parbox{2cm}{\centering DGCF}}
      & BPR   & 0.0423 & 0.1408 & 0.1160  & 0.8860  & 0.9293 & 0.0138 & 0.0328 & 0.0255 & 0.7912 & 0.8346 & 0.2032 & 0.1601 & \cellcolor[HTML]{fed9ae}{0.2522} & 0.6116 & 0.6949  \\
      & +MMR   & 0.0417 & 0.1416 & 0.1169 & \cellcolor[HTML]{fed9ae}{0.8701} & \textbf{\cellcolor[HTML]{fbaf5c}{0.9267}} &  \cellcolor[HTML]{fed9ae}{0.0139} & \cellcolor[HTML]{fed9ae}{0.0331} & \cellcolor[HTML]{fed9ae}{0.0258} & 0.7724 & 0.8276 & 0.2014 & 0.1591 & 0.2471 & 0.5569 & \textbf{\cellcolor[HTML]{fbaf5c}{0.6638}}  \\
      & +Fairness   & \cellcolor[HTML]{fed9ae}{0.0457} & \cellcolor[HTML]{fed9ae}{0.1506} & \cellcolor[HTML]{fed9ae}{0.1228} & 0.8757 & 0.9275 & 0.0121 & 0.0285 & 0.0222 & \cellcolor[HTML]{fed9ae}{0.7593} & \cellcolor[HTML]{fed9ae}{0.8235} & \cellcolor[HTML]{fed9ae}{0.2080} & \cellcolor[HTML]{fed9ae}{0.1670} & 0.2440 & 0.5760 & 0.6779  \\

      & +IPS  & 0.0417 & 0.1392 & 0.1143 & 0.8748 & 0.9279 & 0.0136 & 0.0322 & 0.0251 & 0.7665 & 0.8283 & 0.1995 & 0.1588 & 0.2469 & \cellcolor[HTML]{fed9ae}{0.5562} & \cellcolor[HTML]{fed9ae}{0.6732}  \\
      
      & +CD-CGCN & \textbf{\cellcolor[HTML]{fbaf5c}{0.0470}} & \textbf{\cellcolor[HTML]{fbaf5c}{0.1534}} & \textbf{\cellcolor[HTML]{fbaf5c}{0.1272}} & \textbf{\cellcolor[HTML]{fbaf5c}{0.8620}} & \cellcolor[HTML]{fed9ae}{0.9272} & \textbf{\cellcolor[HTML]{fbaf5c}{0.0162}} & \textbf{\cellcolor[HTML]{fbaf5c}{0.0395}} & \textbf{\cellcolor[HTML]{fbaf5c}{0.0306}} & \textbf{\cellcolor[HTML]{fbaf5c}{0.7496}} & \textbf{\cellcolor[HTML]{fbaf5c}{0.8217}} & \textbf{\cellcolor[HTML]{fbaf5c}{0.2400}} & \textbf{\cellcolor[HTML]{fbaf5c}{0.1867}} & \textbf{\cellcolor[HTML]{fbaf5c}{0.3067}} & \textbf{\cellcolor[HTML]{fbaf5c}{0.5409}} & 0.7049 \\
      \bottomrule
    \end{tabular}}
  \resizebox{\textwidth}{!}{

    \begin{tabular}{|>{\centering\arraybackslash}p{2cm}|p{2.5cm}|>{\centering\arraybackslash}p{1.4cm}>{\centering\arraybackslash}p{1.4cm}>{\centering\arraybackslash}p{1.4cm}>{\centering\arraybackslash}p{1.4cm}>{\centering\arraybackslash}p{1.4cm}| 
               >{\centering\arraybackslash}p{1.4cm}>{\centering\arraybackslash}p{1.4cm}>{\centering\arraybackslash}p{1.4cm}>{\centering\arraybackslash}p{1.4cm}>{\centering\arraybackslash}p{1.4cm}| 
               >{\centering\arraybackslash}p{1.4cm}>{\centering\arraybackslash}p{1.4cm}>{\centering\arraybackslash}p{1.4cm}>{\centering\arraybackslash}p{1.4cm}>{\centering\arraybackslash}p{1.4cm}|}
      \toprule
      \multirow{2}{*}{\parbox{2cm}{\centering \textbf{Models}}} & 
      \multirow{2}{*}{\textbf{Baselines}} & 
      \multicolumn{5}{c|}{\textbf{Gowalla}} & 
      \multicolumn{5}{c|}{\textbf{Amazon-Book}} & 
      \multicolumn{5}{c|}{\textbf{ML-2k}}  \\

        & & \scalebox{0.85}{P@100 $\uparrow$} & \scalebox{0.85}{R@100 $\uparrow$} & \scalebox{0.85}{N@100 $\uparrow$} & \scalebox{0.85}{F@100 $\downarrow$} & \scalebox{0.85}{G@100 $\downarrow$} & 
        \scalebox{0.85}{P@100 $\uparrow$} & \scalebox{0.85}{R@100 $\uparrow$} & \scalebox{0.85}{N@100 $\uparrow$} & \scalebox{0.85}{F@100 $\downarrow$} & \scalebox{0.85}{G@100 $\downarrow$} & 
        \scalebox{0.85}{P@100 $\uparrow$} & \scalebox{0.85}{R@100 $\uparrow$} & \scalebox{0.85}{N@100 $\uparrow$} & \scalebox{0.85}{F@100 $\downarrow$} & \scalebox{0.85}{G@100 $\downarrow$} \\
      \midrule
      \multirow{5}{*}{\parbox{2cm}{\centering MF}} 
      & BPR   & 0.0191 & \cellcolor[HTML]{fed9ae}{0.3108}  & \cellcolor[HTML]{fed9ae}{0.1585}  & 0.8811 & 0.9250 & 0.0089 & 0.1015  & 0.0480 & 0.7498 & 0.8076 & 0.1093 & 0.3838 & \cellcolor[HTML]{fed9ae}{0.2871} & 0.5252 & 0.6343  \\
      & +MMR   & 0.0187 & 0.3061 & 0.1566 & 0.8731 & 0.9230 &  \cellcolor[HTML]{fed9ae}{0.0089} & \cellcolor[HTML]{fed9ae}{0.1017} & \cellcolor[HTML]{fed9ae}{0.0481} & 0.7417 & 0.8041 & 0.1092 & 0.3803 & 0.2778 & 0.5250 & 0.6337  \\
      & +Fairness   & \cellcolor[HTML]{fed9ae}{0.0193} & \textbf{\cellcolor[HTML]{fbaf5c}{0.3120}} & 0.1546 & 0.9005 & 0.9296 & 0.0080 & 0.0905 & 0.0426 & 0.7434 & 0.8058 & \cellcolor[HTML]{fed9ae}{0.1119} & \cellcolor[HTML]{fed9ae}{0.3972} & 0.2744 & 0.5285 & 0.6405  \\

      & +IPS  & 0.0188 & 0.3075 & 0.1549 & \cellcolor[HTML]{fed9ae}{0.8670} & \cellcolor[HTML]{fed9ae}{0.9221} & 0.0078 & 0.0875 & 0.0419 & \cellcolor[HTML]{fed9ae}{0.6989} & \cellcolor[HTML]{fed9ae}{0.7850} & 0.1072 & 0.3726 & 0.2741 & \cellcolor[HTML]{fed9ae}{0.4839} & \cellcolor[HTML]{fed9ae}{0.6200}  \\
      
      & +CD-CGCN & \textbf{\cellcolor[HTML]{fbaf5c}{0.0193}} & 0.3104 & \textbf{\cellcolor[HTML]{fbaf5c}{0.1612}} & \textbf{\cellcolor[HTML]{fbaf5c}{0.8482}} & \textbf{\cellcolor[HTML]{fbaf5c}{0.9190}} & \textbf{\cellcolor[HTML]{fbaf5c}{0.0092}} & \textbf{\cellcolor[HTML]{fbaf5c}{0.1053}} & \textbf{\cellcolor[HTML]{fbaf5c}{0.0514}} & \textbf{\cellcolor[HTML]{fbaf5c}{0.6333}} & \textbf{\cellcolor[HTML]{fbaf5c}{0.7609}} & \textbf{\cellcolor[HTML]{fbaf5c}{0.1155}} & \textbf{\cellcolor[HTML]{fbaf5c}{0.3992}} & \textbf{\cellcolor[HTML]{fbaf5c}{0.3110}} & \textbf{\cellcolor[HTML]{fbaf5c}{0.4568}} & \textbf{\cellcolor[HTML]{fbaf5c}{0.6199}}  \\
      \midrule
      \multirow{5}{*}{\parbox{2cm}{\centering NGCF}}
      & BPR   & 0.0204  & 0.3285 & 0.1713 & 0.8855 & 0.9260 & 0.0095 & 0.1094 & 0.0518 & 0.7603 & 0.8118 & 0.1271 & 0.4406 & 0.3514 & 0.6009 & 0.6851  \\
      & +MMR   & \cellcolor[HTML]{fed9ae}{0.0211} & \cellcolor[HTML]{fed9ae}{0.3433} & \cellcolor[HTML]{fed9ae}{0.1828} & 0.8775 & 0.9244 &  0.0094 & 0.1087 & 0.0514 & 0.7463 & 0.8064 & 0.1280 & \cellcolor[HTML]{fed9ae}{0.4454} & \cellcolor[HTML]{fed9ae}{0.3552} & 0.5908 & 0.6795  \\
      & +Fairness   & 0.0208 & 0.3332 & 0.1723 & 0.9053 & 0.9306 & \cellcolor[HTML]{fed9ae}{0.0097} & \cellcolor[HTML]{fed9ae}{0.1116} & \cellcolor[HTML]{fed9ae}{0.0531} & 0.7585 & 0.8120 & \cellcolor[HTML]{fed9ae}{0.1285} & 0.4431 & 0.3466 & \cellcolor[HTML]{fed9ae}{0.5527} & \textbf{\cellcolor[HTML]{fbaf5c}{0.6623}}  \\

      & +IPS  & 0.0201 & 0.3248 & 0.1695 & \cellcolor[HTML]{fed9ae}{0.8671} & \cellcolor[HTML]{fed9ae}{0.9228} & 0.0091 & 0.1040 & 0.0495 & \cellcolor[HTML]{fed9ae}{0.7175} & \textbf{\cellcolor[HTML]{fbaf5c}{0.7949}} & 0.1231 & 0.4242 & 0.3386 & 0.5436 & \cellcolor[HTML]{fed9ae}{0.6680}  \\
      
      & +CD-CGCN & \textbf{\cellcolor[HTML]{fbaf5c}{0.0220}} & \textbf{\cellcolor[HTML]{fbaf5c}{0.3504}} & \textbf{\cellcolor[HTML]{fbaf5c}{0.1856}} & \textbf{\cellcolor[HTML]{fbaf5c}{0.8317}} & \textbf{\cellcolor[HTML]{fbaf5c}{0.9158}} & \textbf{\cellcolor[HTML]{fbaf5c}{0.0107}} & \textbf{\cellcolor[HTML]{fbaf5c}{0.1239}} & \textbf{\cellcolor[HTML]{fbaf5c}{0.0591}} & \textbf{\cellcolor[HTML]{fbaf5c}{0.7008}} & \cellcolor[HTML]{fed9ae}{0.8025} & \textbf{\cellcolor[HTML]{fbaf5c}{0.1305}} & \textbf{\cellcolor[HTML]{fbaf5c}{0.4520}} & \textbf{\cellcolor[HTML]{fbaf5c}{0.3670}} & \textbf{\cellcolor[HTML]{fbaf5c}{0.5217}} & 0.6744 \\
      \midrule
      \multirow{5}{*}{LightGCN}
      & BPR   & \cellcolor[HTML]{fed9ae}{0.0239} & \cellcolor[HTML]{fed9ae}{0.3777} & \cellcolor[HTML]{fed9ae}{0.2125} & 0.9020 & 0.9303 & \cellcolor[HTML]{fed9ae}{0.0113} & \cellcolor[HTML]{fed9ae}{0.1288} & \cellcolor[HTML]{fed9ae}{0.0623} & 0.7875 & 0.8258 & \cellcolor[HTML]{fed9ae}{0.1313} & \cellcolor[HTML]{fed9ae}{0.4581} & \cellcolor[HTML]{fed9ae}{0.3714} & 0.5902 & 0.6792  \\
      & +MMR   & 0.0225 & 0.3660 & 0.2092 & 0.9030 & 0.9311 &  0.0112 & 0.1285 & 0.0619 & 0.7623 & 0.8161 & 0.1298 & 0.4545 & 0.3680 & 0.5932 & 0.6835  \\
      & +Fairness   & 0.0234 & 0.3685 & 0.2038 & 0.8842 & \cellcolor[HTML]{fed9ae}{0.9266} & 0.0101 & 0.1165 & 0.0552 & 0.7592 & 0.8190 & 0.1269 & 0.4460 & 0.3585 & 0.5750 & \textbf{\cellcolor[HTML]{fbaf5c}{0.6676}}  \\

      & +IPS  & 0.0238 & 0.3767 & 0.2119 & \cellcolor[HTML]{fed9ae}{0.8830} & 0.9277 & 0.0110 & 0.1253 & 0.0606 & \cellcolor[HTML]{fed9ae}{0.7460} & \cellcolor[HTML]{fed9ae}{0.8117} & 0.1309 & 0.4565 & 0.3691 & \cellcolor[HTML]{fed9ae}{0.5602} & \cellcolor[HTML]{fed9ae}{0.6692}  \\
      
      & +CD-CGCN & \textbf{\cellcolor[HTML]{fbaf5c}{0.0241}} & \textbf{\cellcolor[HTML]{fbaf5c}{0.3792}} & \textbf{\cellcolor[HTML]{fbaf5c}{0.2150}} & \textbf{\cellcolor[HTML]{fbaf5c}{0.8767}} & \textbf{\cellcolor[HTML]{fbaf5c}{0.9265}} & \textbf{\cellcolor[HTML]{fbaf5c}{0.0118}} & \textbf{\cellcolor[HTML]{fbaf5c}{0.1350}} & \textbf{\cellcolor[HTML]{fbaf5c}{0.0655}} & \textbf{\cellcolor[HTML]{fbaf5c}{0.7136}} & \textbf{\cellcolor[HTML]{fbaf5c}{0.8025}} & \textbf{\cellcolor[HTML]{fbaf5c}{0.1321}} & \textbf{\cellcolor[HTML]{fbaf5c}{0.4594}} & \textbf{\cellcolor[HTML]{fbaf5c}{0.3753}} & \textbf{\cellcolor[HTML]{fbaf5c}{0.5460}} & 0.6726 \\
      \midrule
      \multirow{5}{*}{\parbox{2cm}{\centering DGCF}}
      & BPR   & 0.0191 & 0.3074 & 0.1654 & 0.8584 & 0.9202 & 0.0090 & 0.1006 & 0.0489 & 0.7728 & 0.8207 & 0.1126 & 0.3950 & 0.3047 & 0.5439 & 0.6446  \\
      & +MMR   & 0.0181 & 0.2929 & 0.1621 & 0.8486 & 0.9180 &  \cellcolor[HTML]{fed9ae}{0.0090} & \cellcolor[HTML]{fed9ae}{0.1010} & \cellcolor[HTML]{fed9ae}{0.0492} & 0.7662 & 0.8182 & 0.1130 & 0.3982 & 0.3019 & 0.5360 & \textbf{\cellcolor[HTML]{fed9ae}{0.6400}}  \\
      & +Fairness   & \cellcolor[HTML]{fed9ae}{0.0207} & \textbf{\cellcolor[HTML]{fbaf5c}{0.3304}} & \cellcolor[HTML]{fed9ae}{0.1762} & \cellcolor[HTML]{fed9ae}{0.8445} & \cellcolor[HTML]{fed9ae}{0.9164} & 0.0080 & 0.0897 & 0.0432 & \cellcolor[HTML]{fed9ae}{0.7296} & \cellcolor[HTML]{fed9ae}{0.8022} & \cellcolor[HTML]{fed9ae}{0.1222} & \cellcolor[HTML]{fed9ae}{0.4265} & \cellcolor[HTML]{fed9ae}{0.3142} & 0.5218 & 0.6406  \\

      & +IPS  & 0.0188 & 0.3033 & 0.1631 & 0.8466 & 0.9178 & 0.0089 & 0.0994 & 0.0483 & 0.7492 & 0.8133 & 0.1117 & 0.3928 & 0.3013 & \cellcolor[HTML]{fed9ae}{0.5048} & \textbf{\cellcolor[HTML]{fbaf5c}{0.6276}}  \\
      
      & +CD-CGCN & \textbf{\cellcolor[HTML]{fbaf5c}{0.0207}} & \cellcolor[HTML]{fed9ae}{0.3255} & \textbf{\cellcolor[HTML]{fbaf5c}{0.1783}} & \textbf{\cellcolor[HTML]{fbaf5c}{0.8276}} & \textbf{\cellcolor[HTML]{fbaf5c}{0.9158}} & \textbf{\cellcolor[HTML]{fbaf5c}{0.0103}} & \textbf{\cellcolor[HTML]{fbaf5c}{0.1174}} & \textbf{\cellcolor[HTML]{fbaf5c}{0.0574}} & \textbf{\cellcolor[HTML]{fbaf5c}{0.7158}} & \textbf{\cellcolor[HTML]{fbaf5c}{0.7996}} & \textbf{\cellcolor[HTML]{fbaf5c}{0.1247}} & \textbf{\cellcolor[HTML]{fbaf5c}{0.4365}} & \textbf{\cellcolor[HTML]{fbaf5c}{0.3517}} & \textbf{\cellcolor[HTML]{fbaf5c}{0.4930}} & 0.6417  \\
      \bottomrule
    \end{tabular}}
\vspace{0.1cm} 
    \begin{tablenotes}
        \footnotesize
        \centering
        \item \colorbox[HTML]{fbaf5c}{\phantom{0}} Best performance \hspace{1cm} \colorbox[HTML]{fed9ae}{\phantom{0}} Suboptimal performance
    \end{tablenotes}

  \label{table:performance_comparison}
\end{table*}
         \subsubsection{Evaluation Metrics.}
         \label{section:metrics}
         We evaluate the overall performance using standard accuracy metrics, along with two innovatively defined filter bubble metrics.
        \begin{itemize}[leftmargin=20pt]
        \item \textbf{Accuracy Metrics.} To evaluate the alignment between the model's recommendation list and the test set, we adopt three standard accuracy metrics: \textbf{Precision@$k$} (\textbf{P@$k$}), \textbf{Recall@$k$} (\textbf{R@$k$}), and \textbf{NDCG@$k$} (\textbf{N@$k$}).

        \label{section:FB_metrics}
        \item \textbf{Filter Bubble Metrics.} To evaluate the degree of filter bubble effect of the model's recommendation list, we design two novel metrics:
            \begin{itemize}
            \item \textbf{Intra-List Filter Bubble Index@$k$} (\textbf{ILFBI@$k$}): The proportion of items in the top-$k$ recommendation list that belong to user's same community. A higher ILFBI indicates a stronger filter bubble effect, and the definition is as follows:
                \begin{equation}
                \textbf{ILFBI}@k=\frac{\sum_{u \in \mathcal{U}}\sum_{i \in \{Top-k\ items\ of\ u\}}\mathbf{1}_{C_i=C_u}}{|\mathcal{U}| \cdot k}
                \label{equation:ILFBI}
                \end{equation}
            \item \textbf{Community Gini Index@$k$} (\textbf{CGI@$k$}): Adapted from economic measures of inequality \cite{GINI}, this index assesses the distribution of items among communities on the recommendation list. A higher CGI reflects a more imbalanced distribution of items across communities in the recommendation list, the definition is as follows:
                \begin{equation}
                \textbf{CGI}@k=1-\frac{2\sum_{i=1}^{n-1}S_i}{nS_n}-\frac{1}{n}
                \label{equation:CGI}
                \end{equation}where $n$ is the total number of communities and $S_i$=$\sum_{j=1}^ic_j$, $[c_1,c_2,...,c_n]$ represents the ascending distribution of the number of communities for the top-$k$ items.            
            \end{itemize}
        
        \end{itemize}


        \subsubsection{Parameter Settings.}
        We implement the CD-CGCN framework in PyTorch and use Adam \cite{adam} as the optimizer. We fix the embedding size as 64 and the global embedding (see details in Sec. \ref{section:CD}) size as 16. The $L_2$ regularization coefficient is $10^{-3}$ for base models (i.e. base embeddings of users and items) and $10^{-7}$ for the Conditional Discriminator Network. For each user we rank all the items (excluding those previously interacted with) and select the highest ranked $k$ items for recommendation, with $k$ chosen from [20, 100]. We strictly split the data into training, validation, and test sets with a ratio of 7:1:2, and community detection is performed exclusively on the training set.

        \subsection{Performance of CD-CGCN (RQ1)}
        \begin{figure}[!h]
          \centering
          \includegraphics[width=1\linewidth]{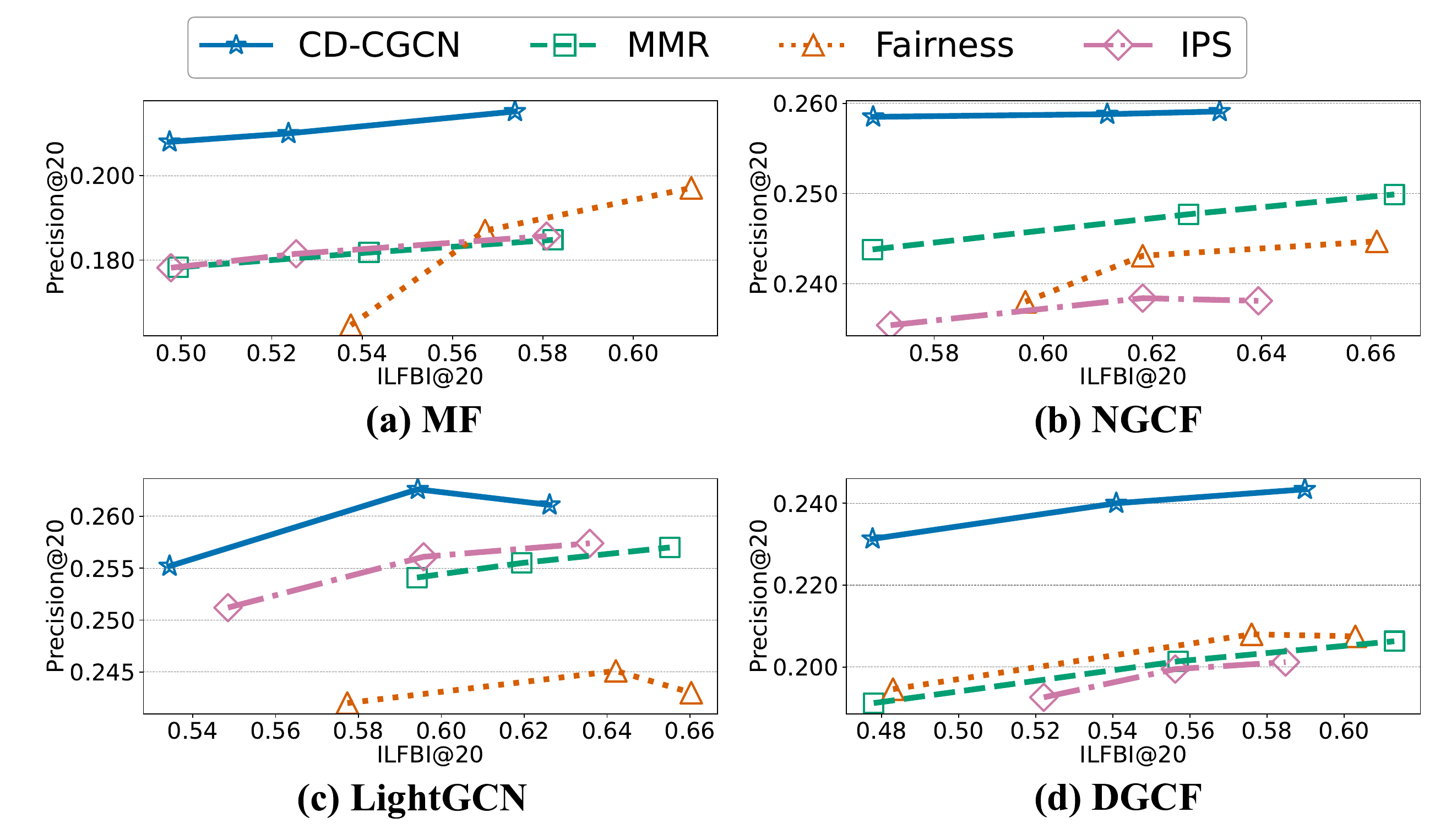}
          \caption{The trade-off curves of \textbf{CD-CGCN}, \textbf{MMR}, \textbf{Fairness} and \text{IPS} across four base models on ML-2k dataset.}
          \label{figure:contrast}
        \end{figure}
        \begin{figure*}[!bh]
          \centering
          \includegraphics[width=1\linewidth]{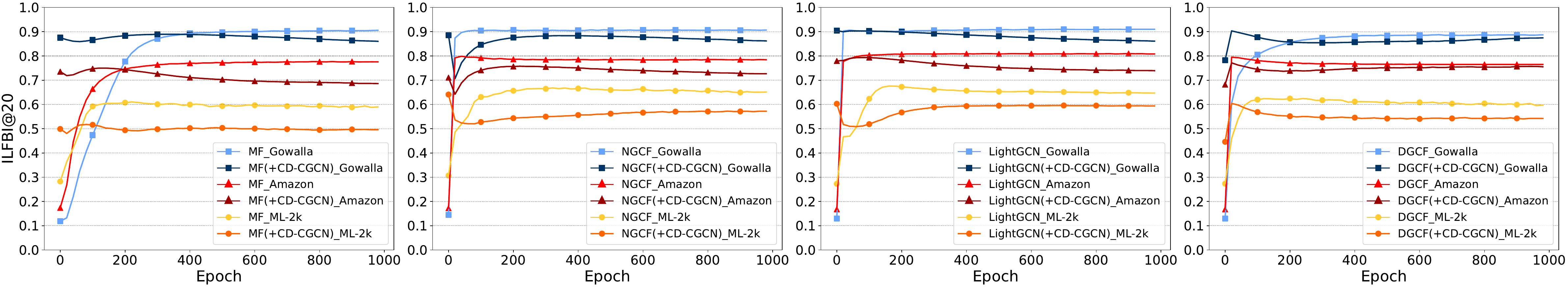}
          \caption{The ILFBI@20 curves \textbf{with and without CD-CGCN} over training epochs.}
          \label{figure:curve_after}
        \end{figure*}
        In this part, we examine how CD-CGCN enhances performance over base models: MF, NGCF, LightGCN and DGCF. The results presented in Table \ref{table:performance_comparison} (with F@$k$ and G@$k$ representing the metrics ILFBI@$k$ and CGI@$k$ mentioned in Sec. \ref{section:metrics}) lead to the following conclusions: 1) CD-CGCN improves performance across all base models on Gowalla, Amazon-Book and ML-2k. 2) Compared with Gowalla and ML-2k, CD-CGCN performs better on Amazon-Book, and we argue that CD-CGCN performs better on larger data sets with more users and items. 3) CD-CGCN shows better performance with NGCF and DGCF than with LightGCN and MF, a possible explanation is that LightGCN and MF are simpler and more lightweight than NGCF and DGCF, and our CD-CGCN achieves greater improvements when integrated with more complex models. 4) We set $k$ to 20 and 100 to better evaluate the performance of CD-CGCN across different recommendation list lengths. As the recommendation list lengthens, improvements in accuracy metrics gradually decrease, whereas improvements in filter bubble metrics increase, which indicates that CD-CGCN mitigates the filter bubble effect more effectively in longer recommendation lists, albeit with some sacrifice in accuracy. 5) Compared with three baselines, CD-CGCN outperforms them significantly in most metrics. Since the post-processing MMR method aims to achieve community balance in the final recommendation list, it has a certain advantage over the CGI metric.

        Since the CD-CGCN and three baselines are all trade-off frameworks, we tune their hyper-parameters on the ML-2k dataset to achieve different accuracy and filter bubble metrics. As illustrated in Fig. \ref{figure:contrast}, the closer the curve is to the upper-left corner, the better the performance in improving accuracy and mitigating filter bubbles. We can also observe that CD-CGCN outperforms three baselines across all base models with different hyper-parameters, which means conditioned on the same level of accuracy, CD-CGCN performs better in mitigating filter bubbles, and vice versa.
        
        In a word, the above experiments demonstrate the superiority of CD-CGCN in mitigating filter bubbles and enhancing accuracy.

        \subsection{Trends of ILFBI during Training and Embedding Visualization (RQ2)}


        \begin{figure}[h]
          \centering
          \includegraphics[width=0.6\linewidth]{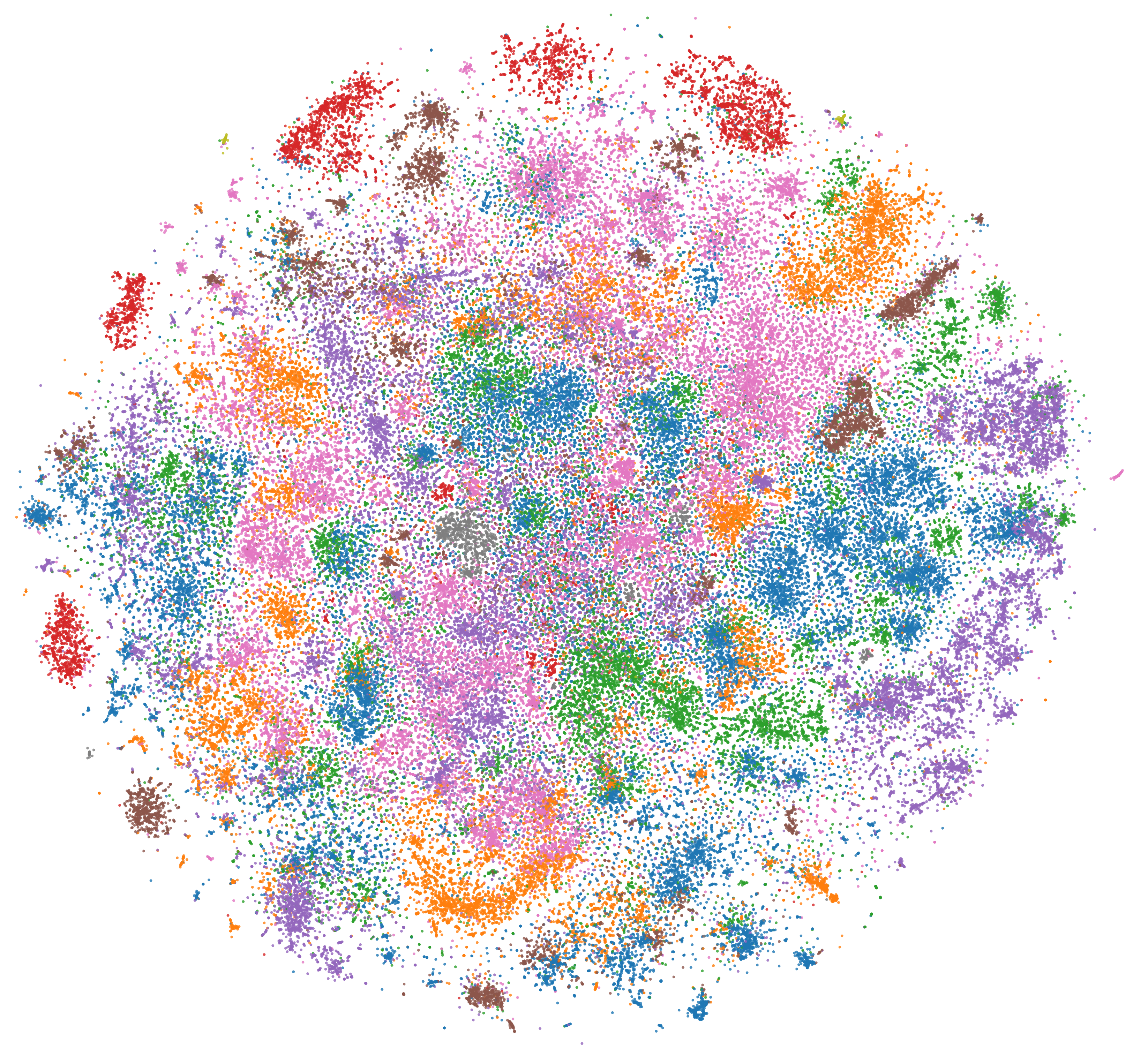}
          \caption{Visualization of embeddings after training LightGCN with \textbf{CD-CGCN} on Amazon-Book dataset, where users and items from the same community are assigned same color.}
          \label{figure:visual_after}
        \end{figure}
        As mentioned in Sec. \ref{section:motivation}, the ILFBI of the recommendation lists from four base models during the training process is much higher than the ILFBI-init. of the training set (see details in Fig. \ref{figure:visual_before}(a)). After applying CD-CGCN, the ILFBI@20 curves (Fig. \ref{figure:curve_after}) remain consistently lower across all base models for all datasets. Initially, the ILFBI values are high with CD-CGCN due to our user-adaptive inference strategy (see details in Sec. \ref{section:user_ada}), which leverages the pretrained best states of each base model and leads to a strong filter bubble effect, but the ILFBI@20 does not increase significantly and remains at a low level during the training process with CD-CGCN. This demonstrates that CD-CGCN enables base models to recommend more inter-community items throughout training process.

        Now let's focus on the role of CD-CGCN in the embeddings of users and items. Reviewing Fig. \ref{figure:visual_before}(b), there is a noticeable clustering of embeddings for users and items within the same community, represented by points of the same color. However, after applying CD-CGCN to LightGCN (see Fig. \ref{figure:visual_after}), we observe that the representations of user and item embeddings within the different communities no longer have distinct boundaries, but instead blend into each other, indicating that users are more likely to be recommended inter-community items, thus alleviating the community multipolarity phenomenon of the filter bubble effect.

        \subsection{Ablation Study (RQ3)}
        \begin{table}[!h]
          \caption{Ablation study on key components of CD-CGCN applied to LightGCN on Amazon-Book dataset.}
          \label{table:ablation}
          \renewcommand{\arraystretch}{1.2}
          \centering
          \scalebox{1}{
          \setlength{\tabcolsep}{1.5mm}{
          \begin{tabular}{c|ccccc}
            \toprule
            Amazon-Book & $\text{P@20}\uparrow$ & $\text{R@20}\uparrow$ & $\text{N@20}\uparrow$ & $\text{F@20}\downarrow$ & $\text{G@20}\downarrow$  \\
            \midrule
            LightGCN &  0.0172  &  0.0416 & 0.0323 & 0.8082 & 0.8402   \\
            w/o CGCN &  0.0175  &  0.0427 & 0.0329 & 0.7634 & 0.8297   \\
            w/o CD & \underline{0.0181} & \underline{0.0441} & \underline{0.0339} & 0.7848 & 0.8338 \\
            w/o CNS &  0.0173  & 0.0419 & 0.0324 & 0.7774 & 0.8349   \\
            w/o UIS &  0.0168  & 0.0402 & 0.0313 & \textbf{0.6243} & \textbf{0.7933}   \\
            \makecell{LightGCN \\ +CD-CGCN} &  \textbf{0.0184}  & \textbf{0.0447} & \textbf{0.0346} & \underline{0.7398} & \underline{0.8228}   \\
            \bottomrule
          \end{tabular}}}
        \end{table}

        To investigate the effectiveness of each component of CD-CGCN, we
        compare it with the following variations based on LightGCN on Amazon-Book.
        1) w/o CGCN removes the Community-reweighted Graph Convolutional Network in Sec. \ref{section:CGCN}.
         2) w/o CD removes the Conditional Discriminator Network in Sec. \ref{section:CD}.  3) w/o CNS changes the Community-enhanced Negative Sampling strategy in Sec. \ref{section:neg_sample} to normal negative sampling strategy. 4) w/o UIS removes the User-adaptive Inference Strategy in Sec. \ref{section:user_ada}, which means that we only use CD-CGCN for inference.

        The results are shown in Table \ref{table:ablation}, it can be observed that: 1) w/o CGCN, w/o CD and w/o CENS have a decline in all metrics, indicating that all three components play a role in improving accuracy and mitigating filter bubble effect. 2) w/o UIS has lower accuracy, suggesting that relying solely on CD-GCGN over-mitigates the filter bubble effect, resulting in significantly reduced accuracy. This highlights the importance of trade-off: rather than excessively sacrificing accuracy to address the filter bubble effect, UIS should be strategically employed to find an optimal trade-off between the accuracy and the filter bubble mitigation.

        \subsection{Effectiveness of CD-CGCN On Debiased Test Set (RQ4)}

        \begin{figure}[!h]
          \centering
          \includegraphics[width=1\linewidth]{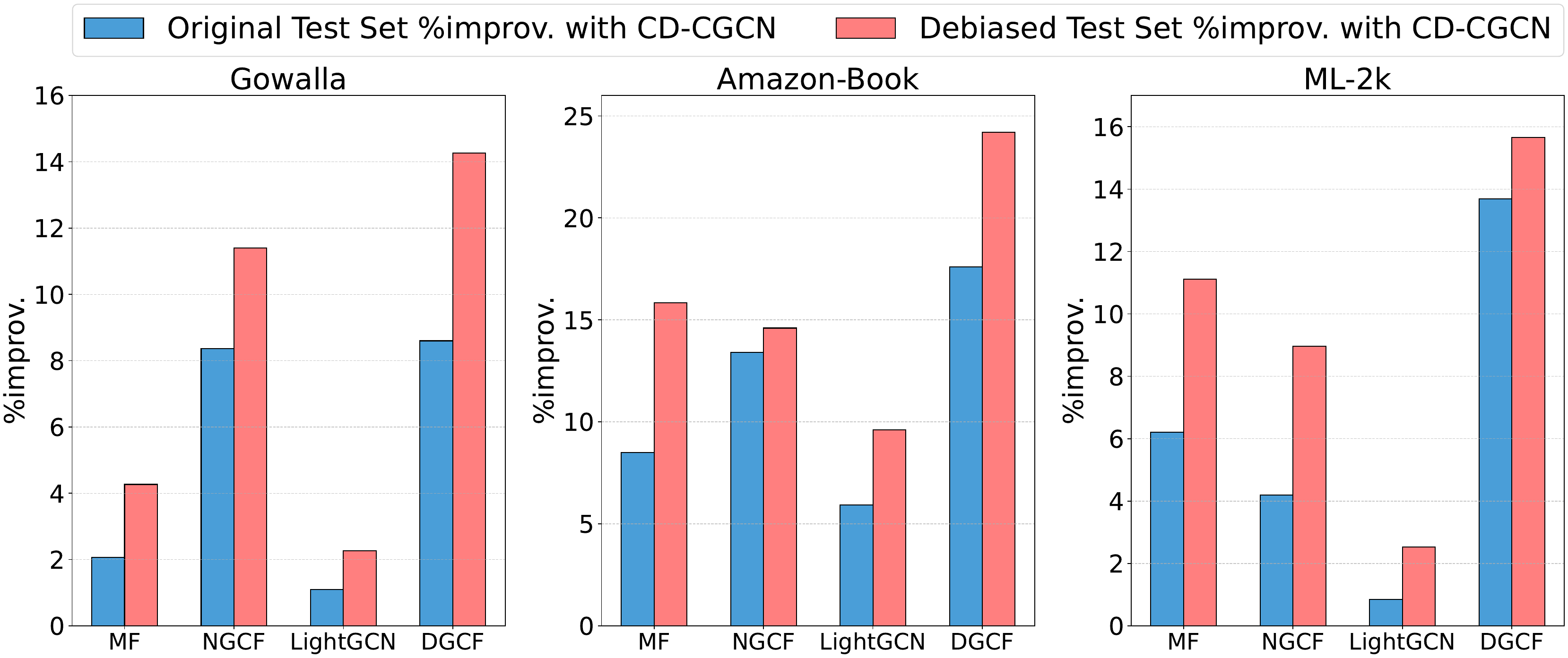}
          \caption{Comparison of CD-CGCN's improvement on original vs. debiased test sets across various base models.}
          \label{figure:debiased}
        \end{figure}
        In this part, we further analyze the original test set of Gowalla and uncover the presence of community bias. Specifically, 80.9\% of the items in the original test set of Gowalla belong to the same community as the corresponding user, indicating significant community bias, and recommending items from the user's own community, as most state-of-the-art recommender systems do, will naturally achieve high accuracy. The situation is similar in Amazon-Book and ML-2k. 

        In order to demonstrate that CD-CGCN can recommend more items of interest to users, with these items coming from different communities, we construct debiased test sets by limiting each user’s test set to a maximum of one item per community. The accuracy on debiased test sets reflects the ability to predict users' preferences across different communities, rather than merely focusing on their own community. Table \ref{table:debiased} illustrates the scale details of the original and the debiased test sets.
        \begin{table}[!h]
          \centering
          \caption{Statistics of original and debiased test sets.}
          \label{table:debiased}
          \scalebox{1}{
          \setlength{\tabcolsep}{0.7mm}{
          \begin{tabular}{c|cc}
            \toprule
            Scale of Data &  Original Test Set  &  Debiased Test Set   \\
            \midrule
            Gowalla &  217,274  &  41,636   \\
            Amazon-Book & 603,378 & 121,829 \\
            ML-2k & 72,701 & 5,655 \\
            \bottomrule
          \end{tabular}}}
        \end{table}
        
        We then re-evaluate the performance of CD-CGCN on these debiased test sets, comparing its accuracy improvements on the debiased test sets with that on the original test sets, as shown in Fig. \ref{figure:debiased} (with the average improvements of P@20, R@20, and N@20 used to measure the accuracy improvement). Our findings indicate that while CD-CGCN shows accuracy improvements on the original test sets, the improvements are significantly greater on the debiased test sets, which proves that CD-CGCN can accurately recommend more inter-community items relevant to users' test sets and predict users' inter-community interests better.

        \subsection{Hyper-parameter Analysis (RQ5)}
            \begin{figure}[!h]
              \centering
              \includegraphics[width=1\linewidth]{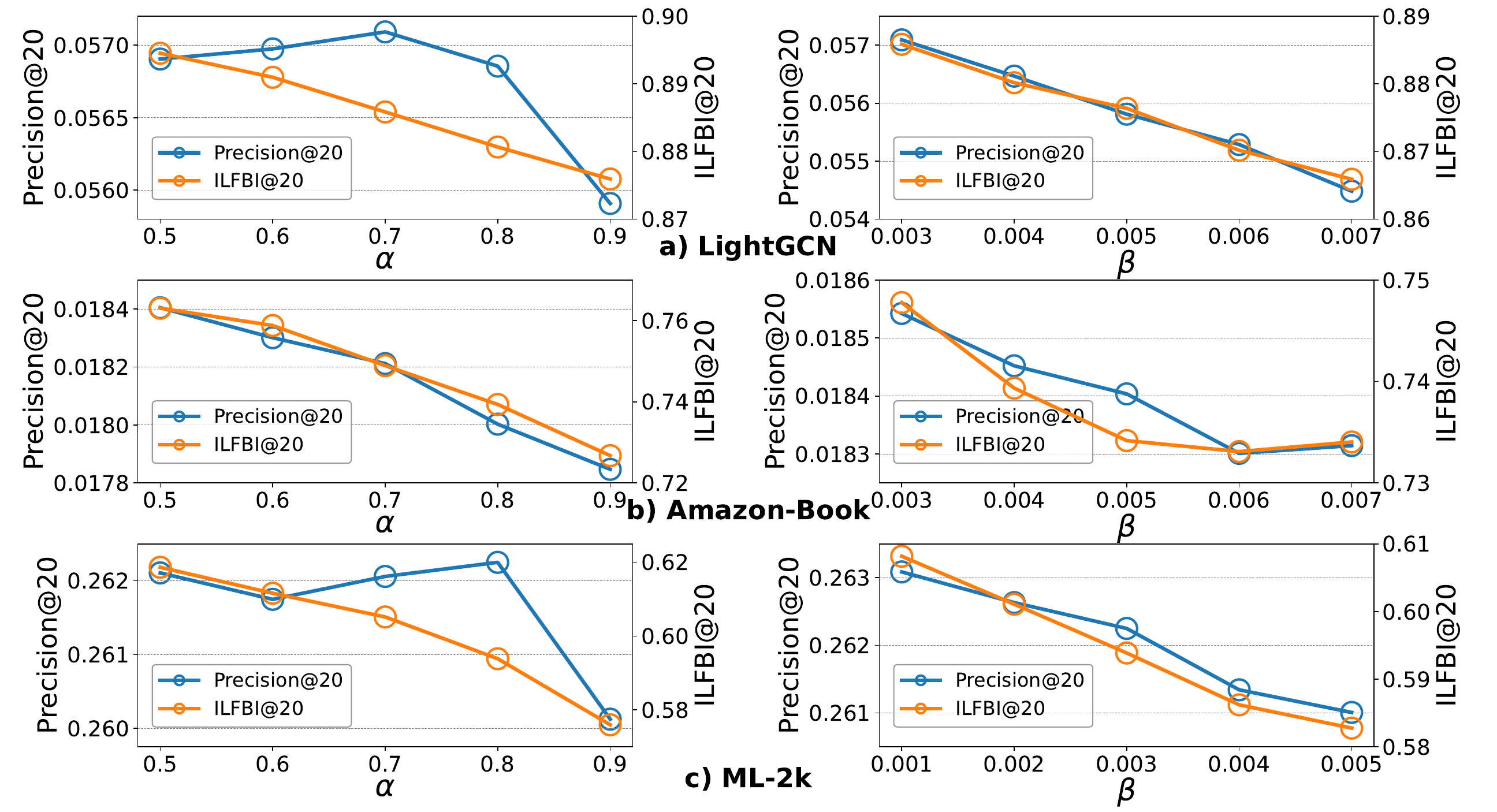}
              \caption{Performance of ILFBI@20 and Precision@20 with different $\alpha$ and $\beta$ based on LightGCN.}
              \label{figure:alpha}
            \end{figure}
        In the proposed CD-CGCN, we introduce two hyper-parameters, $\alpha$ and $\beta$, to control the strength of community-enhanced negative sampling (Sec. \ref{section:neg_sample}) and the weight of the adversarial loss (Sec. \ref{section:beta}), respectively.
        
        As shown in Fig.~\ref{figure:alpha}, increasing $\alpha$ generally leads to a decrease in ILFBI@20 across all datasets, indicating an effective reduction in the filter bubble effect. Meanwhile, Precision@20 exhibits a dataset-dependent trade-off: although it declines overall as $\alpha$ increases, a slight rise in Precision is observed at big $\alpha$ values on the Gowalla and ML-2K datasets. This suggests that moderately introducing community signals during negative sampling can enhance accuracy slightly.
        
        Similarly, as $\beta$ increases, both ILFBI@20 and Precision@20 show a downward trend across all datasets. This indicates that increasing the adversarial loss weight helps mitigate the filter bubble effect, but at the cost of reduced accuracy. 
        
        In a word, by adjusting $\alpha$ and $\beta$, we can flexibly balance the filter bubble effect and the accuracy, similar to the accuracy-diversity trade-off.
\subsection{CASE STUDY}
In this section, we present two case studies on the ML-2k dataset to provide an intuitive understanding of the effectiveness of CD-CGCN from both individual and community perspectives.

    \begin{figure*}[!t]
      \centering
      \includegraphics[width=1\linewidth]{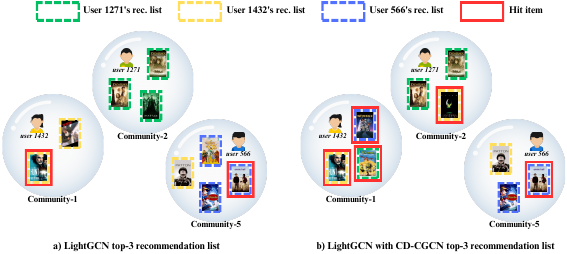}
      \caption{A case study of community perspective on
ML-2k dataset.}
      \label{figure:case2}
    \end{figure*}
    \begin{figure}[!h]
      \centering
      \includegraphics[width=1.0\linewidth]{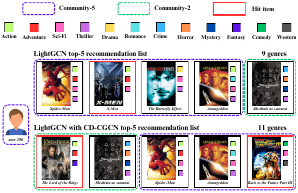}
      \caption{A case study of individual perspective on ML-2k dataset.}
      \label{figure:case}
    \end{figure}
\subsubsection{Case Study from an Individual Perspective}

In this part, we analyze the top-5 recommendation list of the user-296 from an individual perspective, as shown in Fig. \ref{figure:case}. The purple and green boxes highlight the users and items of Community-5 and Community-2, respectively. Under the base model LightGCN, the top-4 recommendation list for the user-296 from Community-5 are all intra-community items, and top-5 items have a total of 9 genres. After incorporating the framework CD-CGCN, the top-2 items in the recommendation list are from Community-2 outside the user-296's own community with the highest-scoring item correctly recommended (highlighted in the red boxes), and the total number of genres increases to 11. This indicates that the CD-CGCN can alleviate the filter bubble effect while capturing user preferences beyond their own community and show a slight improvement in recommendation diversity.

\subsubsection{Case Study from a Community Perspective}


In this part, we analyze the top-3 recommendation lists of three users from different communities from a community perspective, as shown in Fig. \ref{figure:case2}. The items within the green, yellow and blue boxes represent the top-3 recommendation lists of three users, and the users and items within the same bubble indicate that they belong to the same community. Fig. \ref{figure:case2}(a) shows the recommendation result under the base model LightGCN, where the recommended items are mainly from user's own community. In Fig. \ref{figure:case2}(b), after applying CD-CGCN, users' recommendation lists include items from different communities. For example, the recommended items for the user-1432, displayed in the yellow box, come from Community-1, Community-2 and Community-5. Additionally, we observe that the number of correctly recommended items (highlighted in the red boxes) has increased after applying CD-CGCN. This reveals the role of CD-CGCN in alleviating the filter bubble effect from a community perspective.

\section{CONCLUSION AND FUTURE WORK}
In this work, we explore the significant filter bubble effect in existing recommender systems from a novel community-based perspective, highlighting that these recommendations tend to emphasize intra-community items. To address this, we propose a model-agnostic framework CD-CGCN that is compatible with numerous base models to mitigate the filter bubble effect by adversarizing community attributes. We apply CD-CGCN to four base models and conduct extensive experiments on real-world datasets, achieving improvements in both accuracy metrics and our innovatively defined filter bubble metrics compared with other baselines. 

In the future, we will observe the filter bubble phenomenon in multi-round recommendations and use CD-CGCN to verify its effectiveness. Additionally, we plan to further validate the universality of CD-CGCN in other recommendation scenarios, such as explicit feedback rating prediction.
 
\section*{Acknowledgments}
This work was supported in part by the National Key Research and Development Program of China under Grant (2023YFC3310700) and  the National Natural Science Foundation of China (62202041).

\bibliographystyle{IEEEtran}

\bibliography{main}

\begin{thebibliography}{10}
\providecommand{\url}[1]{#1}
\csname url@samestyle\endcsname
\providecommand{\newblock}{\relax}
\providecommand{\bibinfo}[2]{#2}
\providecommand{\BIBentrySTDinterwordspacing}{\spaceskip=0pt\relax}
\providecommand{\BIBentryALTinterwordstretchfactor}{4}
\providecommand{\BIBentryALTinterwordspacing}{\spaceskip=\fontdimen2\font plus
\BIBentryALTinterwordstretchfactor\fontdimen3\font minus
  \fontdimen4\font\relax}
\providecommand{\BIBforeignlanguage}[2]{{%
\expandafter\ifx\csname l@#1\endcsname\relax
\typeout{** WARNING: IEEEtran.bst: No hyphenation pattern has been}%
\typeout{** loaded for the language `#1'. Using the pattern for}%
\typeout{** the default language instead.}%
\else
\language=\csname l@#1\endcsname
\fi
#2}}
\providecommand{\BIBdecl}{\relax}
\BIBdecl

\bibitem{01info}
B.~M. Sarwar, G.~Karypis, J.~A. Konstan, and J.~Riedl, ``Item-based
  collaborative filtering recommendation algorithms,'' in \emph{Proceedings of
  the 10th international conference on World Wide Web}, 2001, p. 285–295.

\bibitem{social_network}
K.~Duskin, J.~Schafer, J.~West, and E.~Spiro, ``Echo chambers in the age of
  algorithms: An audit of twitter’s friend recommender system,'' in
  \emph{16th ACM Web Science Conference 2024 (WebSci ’24)}, 2024.

\bibitem{giu2016online_video}
G.~Fontanini, M.~Bertini, and A.~D. Bimbo, ``Web video popularity prediction
  using sentiment and content visual features,'' in \emph{Proceedings of the
  2016 ACM on international conference on multimedia retrieval}, 2016, p.
  289–292.

\bibitem{wu2019online_news}
C.~Wu, F.~Wu, M.~An, J.~Huang, Y.~Huang, and X.~Xie, ``Npa: Neural news
  recommendation with personalized attention,'' in \emph{Proceedings of the
  25th ACM SIGKDD International Conference on Knowledge Discovery \& Data
  Mining}, 2019, p. 2576–2584.

\bibitem{wang18e-commerce}
J.~Wang, P.~Huang, H.~Zhao, Z.~Zhang, B.~Zhao, and D.~L. Lee, ``Billion-scale
  commodity embedding for e-commerce recommendation in alibaba,'' in
  \emph{Proceedings of the 24th ACM SIGKDD International Conference on
  Knowledge Discovery \& Data Mining}, 2018, pp. 839--848.

\bibitem{17NCF}
X.~He, L.~Liao, H.~Zhang, L.~Nie, X.~Hu, and T.-S. Chua, ``Neural collaborative
  filtering,'' in \emph{Proceedings of the 26th international conference on
  World Wide Web}, 2017, p. 173–182.

\bibitem{14FB}
T.~T. Nguyen, P.-M. Hui, F.~M. Harper, L.~Terveen, and J.~A. Konstan,
  ``Exploring the filter bubble: The effect of using recommender systems on
  content diversity,'' in \emph{Proceedings of the 23rd International
  Conference on World Wide Web}, 2014, p. 677–686.

\bibitem{21FB}
P.~Liu, K.~Shivaram, A.~Culotta, M.~A. Shapiro, and M.~Bilgic, ``The
  interaction between political typology and filter bubbles in news
  recommendation algorithms,'' in \emph{Proceedings of the Proceedings of the
  Web Conference}, 2021, p. 3791–3801.

\bibitem{McKay}
D.~McKay, K.~Owyong, S.~Makri, and M.~G. Lopez, ``Turn and face the strange:
  investigating filter bubble bursting information interactions,'' in
  \emph{Proceedings of the 2022 Conference on Human Information Interaction and
  Retrieval}, 2022, p. 233–242.

\bibitem{Pa11FB}
E.~Pariser, ``The filter bubble: What the internet is hiding from you. london:
  Penguin.'' 2011.

\bibitem{23netpo}
Interian, R., Marzo, R.~G., Mendoza, I., Ribeiro, and C.~C., ``Network
  polarization, filter bubbles, and echo chambers: An annotated review of
  measures and reduction methods,'' in \emph{International Transactions in
  Operational Research, 30(6)}, 2023, pp. 3122--3158.

\bibitem{media}
L.~Terren and R.~Borge, ``Echo chambers on social media: A systematic review of
  the literature,'' in \emph{Review of Communication Research, 9}, 2021, p.
  99–118.

\bibitem{drake16fb}
D.~Baer, ``The ‘filter bubble’ explains why trump won and you didn't see it
  coming. ny mag.'' 2016.

\bibitem{under_EC}
Y.~Ge, S.~Zhao, H.~Zhou, C.~Pei, F.~Sun, and W.~Ou, ``Understanding echo
  chambers in e-commerce eecommender systems,'' in \emph{Proceedings of the
  43rd International ACM SIGIR Conference on Research and Development in
  Information Retrieval}, 2020, p. 2261–2270.

\bibitem{gao22miFB}
Z.~Gao, T.~Shen, Z.~Mai, M.~R. Bouadjenek, I.~Waller, A.~Anderson, R.~Bodkin,
  and S.~Sanner, ``Mitigating the filter bubble while maintaining relevance:
  Targeted diversification with vae-based recommender systems,'' in
  \emph{Proceedings of the 45th International ACM SIGIR Conference on Research
  and Development in Information Retrieval}, 2022, p. 2524–2531.

\bibitem{wang22FB}
W.~Wang, F.~Feng, L.~Nie, and T.-S. Chua, ``User-controllable recommendation
  against filter bubbles,'' in \emph{Proceedings of the 45th International ACM
  SIGIR Conference on Research and Development in Information Retrieval}, 2022,
  p. 1251–1261.

\bibitem{22mifb}
V.~Anand, M.~Yang, and Z.~Zhao, ``Mitigating filter bubbles within deep
  recommender systems. arxiv:2209.08180.'' 2022.

\bibitem{08louvain}
V.~D. Blondel, J.-L. Guillaume, R.~Lambiotte, and E.~Lefebvre, ``Fast unfolding
  of communities in large networks,'' in \emph{Journal of Statistical
  Mechanics: Theory and Experiment 10 (2008)}, 2008.

\bibitem{19loop}
W.~Sun, S.~Khenissi, O.~Nasraoui, and P.~Shafto, ``Debiasing the
  human-recommender system feedback loop in collaborative fltering,'' in
  \emph{Companion Proceedings of The 2019 World Wide Web Conference}, 2019, p.
  645–651.

\bibitem{20loop}
M.~Mansoury, H.~Abdollahpouri, M.~Pechenizkiy, B.~Mobasher, and R.~Burke,
  ``Feedback loop and bias amplifcation in recommender systems,'' in
  \emph{Proceedings of the 29th ACM international conference on information \&
  knowledge management}, 2020, p. 2145–2148.

\bibitem{po_so}
U.~Chitra and C.~Musco, ``Understanding filter bubbles and polarization in
  social networks. arxiv:1906.08772.'' 2019.

\bibitem{gowalla}
D.~Liang, L.~Charlin, J.~McInerney, and D.~M. Blei, ``Modeling user exposure in
  recommendation,'' in \emph{Proceedings of the 25th international conference
  on World Wide Web}, 2016, pp. 951--961.

\bibitem{lightgcn}
X.~He, K.~Deng, X.~Wang, Y.~Li, Y.~Zhang, and M.~Wang, ``Lightgcn: Simplifying
  and powering graph convolution network for recommendation,'' in
  \emph{Proceedings of the 43rd International ACM SIGIR conference on research
  and development in Information Retrieval}, 2020, pp. 639--648.

\bibitem{NGCF}
X.~Wang, X.~He, M.~Wang, F.~Feng, and T.-S. Chua, ``Neural graph collaborative
  filtering,'' in \emph{Proceedings of the 42nd international ACM SIGIR
  conference on Research and development in Information Retrieval}, 2019, p.
  165–174.

\bibitem{MF}
Y.~Koren, R.~Bell, and C.~Volinsky, ``Matrix factorization techniques for
  recommender systems,'' in \emph{Computer 42, 8 (2009)}, 2009, pp. 30--37.

\bibitem{DGCF}
X.~Wang, H.~Jin, A.~Zhang, X.~He, T.~Xu, and T.-S. Chua, ``Disentangled graph
  collaborative filtering,'' in \emph{Proceedings of the 43rd international ACM
  SIGIR conference on research and development in information retrieval}, 2020,
  p. 1001–1010.

\bibitem{amazon}
R.~He and J.~McAuley, ``Ups and downs: Modeling the visual evolution of fashion
  trends with one-class collaborative filtering,'' in \emph{proceedings of the
  25th international conference on world wide web}, 2016, pp. 507--517.

\bibitem{tsne}
L.~V. der Maaten and G.~Hinton, ``Visualizing data using t-sne,'' in
  \emph{Journal of machine learning research 9, 11}, 2008.

\bibitem{bpr}
S.~Rendle, C.~Freudenthaler, Z.~Gantner, and L.~Schmidt-Thieme, ``Bpr: Bayesian
  personalized ranking from implicit feedback,'' in \emph{Proceedings of the
  Twenty-Fifth Conference on Uncertainty in Artificial Intelligence}, 2009.

\bibitem{CELoss}
Shannon and C.~Elwood., ``A mathematical theory of communication,'' in
  \emph{The Bell system technical journal}, 1948, pp. 379--423.

\bibitem{17GCN}
T.~N. Kipf and M.~Welling, ``Semi-supervised classification with graph
  convolutional networks,'' in \emph{International Conference on Learning
  Representations (ICLR)}, 2017.

\bibitem{liu2020loss}
C.~Liu, M.~Salzmann, T.~Lin, R.~Tomioka, and S.~S{\"u}sstrunk, ``On the loss
  landscape of adversarial training: Identifying challenges and how to overcome
  them,'' in \emph{Advances in Neural Information Processing Systems}, 2020,
  pp. 21\,476--21\,487.

\bibitem{GRL}
Y.~Ganin and V.~S. Lempitsky, ``Unsupervised domain adaptation by
  backpropagation,'' in \emph{Proceedings of the 32nd International Conference
  on Machine Learning (Proceedings of Machine Learning Research)}, 2015, pp.
  1180--1189.

\bibitem{Cantador:RecSys2011}
K.~T. Cantador~I, Brusilovsky~P, ``2nd workshop on information heterogeneity
  and fusion in recommender systems (hetrec 2011),'' in \emph{Proceedings of
  the 5th ACM conference on Recommender systems}, 2011, pp. 387--388.

\bibitem{98Carbonell}
C.~J and G.~J., ``The use of mmr, diversity-based reranking for reordering
  documents and producing summaries,'' in \emph{Proceedings of the 21st annual
  international ACM SIGIR conference on Research and development in information
  retrieval}, 1998, pp. 335--336.

\bibitem{19Beutel}
B.~A, C.~J, and D.~T., ``Fairness in recommendation ranking through pairwise
  comparisons,'' in \emph{Proceedings of the 25th ACM SIGKDD international
  conference on knowledge discovery \& data mining}, 2019, pp. 2212--2220.

\bibitem{joachims2017unbiased}
T.~Joachims, A.~Swaminathan, and T.~Schnabel, ``Unbiased learning-to-rank with
  biased feedback,'' in \emph{Proceedings of the tenth ACM international
  conference on web search and data mining}, 2017, pp. 781--789.

\bibitem{schnabel2016recommendations}
T.~Schnabel, A.~Swaminathan, A.~Singh, N.~Chandak, and T.~Joachims,
  ``Recommendations as treatments: Debiasing learning and evaluation,'' in
  \emph{international conference on machine learning}.\hskip 1em plus 0.5em
  minus 0.4em\relax PMLR, 2016, pp. 1670--1679.

\bibitem{GINI}
A.~Antikacioglu and R.~Ravi, ``Post processing recommender systems for
  diversity,'' in \emph{Proceedings of the 23rd ACM SIGKDD International
  Conference on Knowledge Discovery and Data Mining}, 2017, p. 707–716.

\bibitem{adam}
D.~P. Kingma and J.~Ba, ``Adam: A method for stochastic optimization,'' in
  \emph{3rd International Conference on Learning Representations, ICLR}, 2015.

\bibitem{pares2018fluid}
F.~Par{\'e}s, D.~G. Gasulla, A.~Vilalta, J.~Moreno, E.~Ayguad{\'e}, J.~Labarta,
  U.~Cort{\'e}s, and T.~Suzumura, ``Fluid communities: A competitive, scalable
  and diverse community detection algorithm,'' in \emph{Complex Networks \&
  Their Applications VI: Proceedings of Complex Networks 2017 (The Sixth
  International Conference on Complex Networks and Their Applications)}.\hskip
  1em plus 0.5em minus 0.4em\relax Springer, 2018, pp. 229--240.

\bibitem{zhang2018understanding}
Y.~Zhang and K.~Rohe, ``Understanding regularized spectral clustering via graph
  conductance,'' \emph{Advances in Neural Information Processing Systems},
  vol.~31, 2018.

\bibitem{reichardt2006statistical}
J.~Reichardt and S.~Bornholdt, ``Statistical mechanics of community
  detection,'' \emph{Physical Review E—Statistical, Nonlinear, and Soft
  Matter Physics}, vol.~74, no.~1, p. 016110, 2006.

\bibitem{traag2019louvain}
V.~A. Traag, L.~Waltman, and N.~J. Van~Eck, ``From louvain to leiden:
  guaranteeing well-connected communities,'' \emph{Scientific reports}, vol.~9,
  no.~1, pp. 1--12, 2019.

\bibitem{leicht2008community}
E.~A. Leicht and M.~E. Newman, ``Community structure in directed networks,''
  \emph{Physical review letters}, vol. 100, no.~11, p. 118703, 2008.

\bibitem{radicchi2004defining}
F.~Radicchi, C.~Castellano, F.~Cecconi, V.~Loreto, and D.~Parisi, ``Defining
  and identifying communities in networks,'' \emph{Proceedings of the national
  academy of sciences}, vol. 101, no.~9, pp. 2658--2663, 2004.

\bibitem{yang2012defining}
J.~Yang and J.~Leskovec, ``Defining and evaluating network communities based on
  ground-truth,'' in \emph{Proceedings of the ACM SIGKDD workshop on mining
  data semantics}, 2012, pp. 1--8.

\bibitem{fortunato2010community}
S.~Fortunato, ``Community detection in graphs,'' \emph{Physics reports}, vol.
  486, no. 3-5, pp. 75--174, 2010.

\bibitem{shi2000normalized}
J.~Shi and J.~Malik, ``Normalized cuts and image segmentation,'' \emph{IEEE
  Transactions on pattern analysis and machine intelligence}, vol.~22, no.~8,
  pp. 888--905, 2000.

\bibitem{flake2000efficient}
G.~W. Flake, S.~Lawrence, and C.~L. Giles, ``Efficient identification of web
  communities,'' in \emph{Proceedings of the sixth ACM SIGKDD international
  conference on Knowledge discovery and data mining}, 2000, pp. 150--160.

\bibitem{newman2004finding}
M.~E. Newman and M.~Girvan, ``Finding and evaluating community structure in
  networks,'' \emph{Physical review E}, vol.~69, no.~2, p. 026113, 2004.

\bibitem{zhang2010determining}
S.~Zhang, X.-M. Ning, C.~Ding, and X.-S. Zhang, ``Determining modular
  organization of protein interaction networks by maximizing modularity
  density,'' in \emph{BMC systems biology}, vol.~4.\hskip 1em plus 0.5em minus
  0.4em\relax Springer, 2010, pp. 1--12.

\bibitem{newsFB}
L.~Michiels, J.~Vannieuwenhuyze, J.~Leysen, R.~Verachtert, A.~Smets, and
  B.~Goethals, ``How should we measure filter bubbles? a regression model and
  evidence for online news,'' in \emph{Proceedings of the 17th ACM Conference
  on Recommender Systems (RecSys ’23)}, 2023, p. 640–651.

\bibitem{FBvis}
S.~Nagulendra and J.~Vassileva, ``Understanding and controlling the filter
  bubble through interactive visualization: a user study,'' in
  \emph{Proceedings of the 25th ACM conference on Hypertext and social media},
  2014, p. 107–115.

\bibitem{cinelli}
M.~Cinelli, G.~D.~F. Morales, A.~Galeazzi, W.~Quattrociocchi, and M.~Starnini,
  ``The echo chamber eefect on social media,'' in \emph{Proceedings of the
  National Academy of Sciences}, 2021.

\bibitem{aridor}
G.~Aridor, D.~Goncalves, and S.~Sikdar, ``Deconstructing the filter bubble:
  User decision-making and recommender systems,'' in \emph{Proceedings of the
  14th ACM Conference on Recommender Systems}, 2020, p. 82–91.

\bibitem{22nianli}
N.~Li, C.~Gao, J.~Piao, X.~Huang, A.~Yue, L.~Zhou, Q.~Liao, and Y.~Li, ``An
  exploratory study of information cocoon on shortform video platform,'' in
  \emph{Proceedings of the 31st ACM International Conference on Information \&
  Knowledge Management}, 2022, p. 4178–4182.

\bibitem{23piao}
J.~Piao, J.~Liu, F.~Zhang, J.~Su, and Y.~Li, ``Human–ai adaptive dynamics
  drives the emergence of information cocoons,'' in \emph{Nature Machine
  Intelligence}, 2023, pp. 1--11.

\bibitem{20Chitra}
U.~Chitra and C.~Musco, ``Analyzing the impact of filter bubbles on social
  network polarization,'' in \emph{Proceedings of the 13th International
  Conference on Web Search and Data Mining}, 2020, p. 115–123.

\bibitem{23Zhang}
Z.~H, Z.~Z, and C.~J., ``Evolution of filter bubbles and polarization in news
  recommendation,'' in \emph{European Conference on Information Retrieval.
  Cham: Springer Nature Switzerland}, 2023, pp. 685--693.

\bibitem{CouFB}
Symeonidis, Panagiotis, L.~Coba, and M.~Zanker, ``Counteracting the filter
  bubble in recommender systems: Novelty-aware matrix factorization,'' in
  \emph{Intelligenza Artificiale 13 (1)}, 2019, pp. 37--47.

\bibitem{PFB}
J.~Sun, J.~Song, Y.~Jiang, Y.~Liu, and J.~Li, ``Prick the flter bubble: A novel
  cross domain recommendation model with adaptive diversity regularization,''
  in \emph{Electronic Markets.}, 2021.

\bibitem{grossetti2019community}
Q.~Grossetti, C.~Du~Mouza, and N.~Travers, ``Community-based recommendations on
  twitter: Avoiding the filter bubble,'' in \emph{Web Information Systems
  Engineering--WISE 2019: 20th International Conference, Hong Kong, China,
  January 19--22, 2020, Proceedings 20}.\hskip 1em plus 0.5em minus 0.4em\relax
  Springer, 2019, pp. 212--227.

\bibitem{Gao}
C.~Gao, S.~Wang, S.~Li, J.~Chen, X.~He, and W.~Lei, ``Cirs: Bursting filter
  bubbles by counterfactual interactive recommender system,'' in \emph{ACM
  Transactions on Information Systems}, 2023, pp. 1--27.

\bibitem{23Xu}
Y.~Xu, E.~Wang, Y.~Yang, and H.~Xiong, ``Gs-rs: A generative approach for
  alleviating cold start and filter bubbles in recommender systems,'' in
  \emph{IEEE Transactions on Knowledge and Data Engineering}, 2024, pp.
  668--681.

\bibitem{20masrour}
F.~Masrour, T.~Wilson, H.~Yan, P.~Tan, and A.~Esfahanian, ``Bursting the filter
  bubble: Fairness-aware network link prediction,'' in \emph{Proceedings of the
  AAAI conference on artificial intelligence}, 2020, p. 841–848.

\bibitem{21Tommasel}
A.~Tommasel, J.~M. Rodriguez, and D.~Godoy, ``I want to break free!
  recommending friends from outside the echo chamber,'' in \emph{Proceedings of
  the 15th ACM Conference on Recommender Systems}, 2021, p. 23–33.

\bibitem{li2023breaking}
Z.~Li, Y.~Dong, C.~Gao, Y.~Zhao, D.~Li, J.~Hao, K.~Zhang, Y.~Li, and Z.~Wang,
  ``Breaking filter bubble: A reinforcement learning framework of controllable
  recommender system,'' in \emph{Proceedings of the ACM Web Conference 2023},
  2023, pp. 4041--4049.

\end{thebibliography}
\balance

\newpage
\appendix

\subsection{Equivalence Proof}
\label{App}
Here we prove that under the definition of the GRL (see Equation \ref{equation:12}), Equation \ref{equation:15} is equivalent to Equations \ref{equation:10} and \ref{equation:11}.
First by Equations \ref{equation:9}, \ref{equation:10} and \ref{equation:11}, we have parameters updated in the following way:
    
    \begin{align}
    \theta_b &\leftarrow\theta_b-\mu_1 (\frac{\partial \mathcal{L}_{rec}(\theta_b)}{\partial \theta_b} - \beta\frac{\partial \mathcal{L}_{adv}(\theta_b,\theta_d)}{\partial \theta_b}) \nonumber \\
    &= \theta_b-\mu_1 \frac{\partial \mathcal{L}_{rec}(\theta_b)}{\partial \theta_b} +\mu_1 \beta\frac{\partial \mathcal{L}_{adv}(\theta_b,\theta_d)}{\partial \theta_b}
    \label{equation:17}
    \end{align}

    \begin{equation}
    \begin{aligned}
    \theta_d\leftarrow\theta_d-\mu_2\beta\frac{\partial \mathcal{L}_{adv}(\theta_b,\theta_d)}{\partial \theta_d}  \\ 
    \label{equation:18}
    \end{aligned}
    \end{equation}where $\mu_1$ and $\mu_2$ are the learning rates of $\theta_b$ and $\theta_d$, respectively.

Next, we derive the parameter update method from Equations \ref{equation:14} and \ref{equation:15} under the definition of the GRL:

\begin{align}
    \theta_b &\leftarrow \theta_b - \mu \left( \frac{\partial \mathcal{L}_{rec}(\theta_b)}{\partial \theta_b} + \frac{\partial \mathcal{L}_{adv}(\mathcal{R}_\beta(\theta_b), \theta_d)}{\partial \theta_b} \right) \nonumber \\
    &= \theta_b - \mu \left( \frac{\partial \mathcal{L}_{rec}(\theta_b)}{\partial \theta_b} + \frac{\partial \mathcal{L}_{adv}(\mathcal{R}_\beta(\theta_b), \theta_d)}{\partial \mathcal{R}_\beta(\theta_b)} \frac{\partial \mathcal{R}_\beta(\theta_b)}{\partial \theta_b} \right)\nonumber \\
    &= \theta_b - \mu \left( \frac{\partial \mathcal{L}_{rec}(\theta_b)}{\partial \theta_b} + \frac{\partial \mathcal{L}_{adv}(\theta_b, \theta_d)}{\partial \theta_b} (-\beta) \right) \nonumber \\
    &= \theta_b - \mu  \frac{\partial \mathcal{L}_{rec}(\theta_b)}{\partial \theta_b} + \mu \beta\frac{\partial \mathcal{L}_{adv}(\theta_b, \theta_d)}{\partial \theta_b}  
    \label{equation:19}
\end{align}

\begin{align}
    \theta_d &\leftarrow \theta_d - \mu \frac{\partial \mathcal{L}_{adv}(\mathcal{R}_\beta(\theta_b), \theta_d)}{\partial \theta_d}  \nonumber \\
    &=\theta_d - \mu \frac{\partial \mathcal{L}_{adv}(\theta_b, \theta_d)}{\partial \theta_d} 
    \label{equation:20}
\end{align}where $\mu$ is the learning rate.

By comparing Equation \ref{equation:17} to Equation \ref{equation:19}, as well as Equation \ref{equation:18} to Equation \ref{equation:20}, we find that their forms are identical, which confirms the validation described above.

\begin{table*}[!b]
\centering
\caption{Comparison of community detection algorithms across multiple metrics on the Gowalla dataset partitioned into 18 communities.}
\label{table:community detection}
\renewcommand{\arraystretch}{1.1}
\resizebox{\textwidth}{!}{
\begin{tabular}{c|cccccccccccc}
\toprule
Algorithms & IED $\uparrow$ & EI $\uparrow$ & AID $\uparrow$ & FOMD $\uparrow$ & E $\downarrow$ & CR $\downarrow$ & C $\downarrow$ & NC $\downarrow$ & MODF $\downarrow$ & FODF $\downarrow$ & NGM $\uparrow$ & MD $\uparrow$ \\
\midrule
FluidC       & 0.0042 & 32568 & 15.78 & \underline{0.4695} & 6.425 & 9.597e-5 & 0.2891 & 0.3049 & 495.1 & 0.1476 & 0.6564 & 168.4 \\
R-spectral & 0.0076 & 35721 & 16.25 & \textbf{0.4708} & 4.438 & 6.673e-5 & 0.2140 & 0.2261 & 358.2 & 0.1342 & 0.6901 & 212.7 \\
Spin-glass   & \underline{0.0109} & 36008 & 15.54 & 0.4654 & 4.563 & 6.832e-5 & 0.2322 & 0.2439 & 406.8 & 0.0983 & 0.6990 & 197.7 \\
Leiden      & 0.0105 & \underline{37860} & \underline{16.40} & 0.4685 & 3.672 & \underline{5.499e-5} & 0.1822 & 0.1916 & \underline{341.2} & \underline{0.0624} & \underline{0.7182} & \underline{229.1} \\
Rb-pots    & 0.0093 & 36941 & \textbf{16.47} & 0.4681 & \underline{3.666} & 5.526e-5 & \underline{0.1808} & \underline{0.1914} & 366.8 & \textbf{0.0605} & \textbf{0.7190} & \textbf{230.4} \\
Louvain     & \textbf{0.0134} & \textbf{38132} & 15.99 & 0.4664 & \textbf{3.510} & \textbf{5.267e-5} & \textbf{0.1808} & \textbf{0.1900} & \textbf{262.6} & 0.0658 & 0.7050 & 224.7 \\
\bottomrule
\end{tabular}
}
\end{table*}
\subsection{Performance Comparison of Louvain and Other Community Detection Algorithms}
To validate the rationality of employing the Louvain \cite{08louvain} algorithm for community detection in our main experiments, we conduct a comparative evaluation against several representative algorithms.
        \begin{itemize}[leftmargin=20pt]
        \item \textbf{Compared Algorithms:} We employ several well-known community detection algorithms for comparison as follows: FluidC \cite{pares2018fluid}, R-spectral \cite{zhang2018understanding}, Spin-lass \cite{reichardt2006statistical}, Leiden \cite{traag2019louvain} and Rb-pots \cite{leicht2008community}.
        \item \textbf{Quality Metrics:} We assess the performance of these community detection algorithms across four classes: 1) Internal connectivity metrics: Internal Edge Density (IED) \cite{radicchi2004defining}, Edges Inside (EI) \cite{radicchi2004defining}, Average Internal Degree (AID) \cite{radicchi2004defining}, Fraction Over Median Degree (FOMD) \cite{yang2012defining}. 2) External connectivity metrics: Expansion (E) \cite{yang2012defining}, Cut Ratio (CR) \cite{fortunato2010community}. 3) Combined internal-external metrics: Conductance (C) \cite{shi2000normalized}, Normalized Cut (NC) \cite{shi2000normalized}, Maximum Out Degree Fraction (MODF) \cite{flake2000efficient}, Flake Out Degree Fraction (FODF) \cite{flake2000efficient}. 4) Modularity-based metrics: Newman Girvan Modularity (NGM) \cite{newman2004finding}, Modularity Density (MD) \cite{zhang2010determining}.
        \end{itemize}

As shown in Table \ref{table:community detection}, the Louvain algorithm achieves the best performance on most evaluation metrics, and its performance on the remaining metrics is also competitive. These results confirm the feasibility and appropriateness of using the Louvain algorithm for community detection in our main text.

\subsection{Related Work}

\subsubsection{Discovery and Definition of Filter Bubble Effect in Recommender Systems}
The filter bubble effect in recommender systems was first introduced by Nguyen \textit{et al.} \cite{14FB}, highlighting how the diversity of recommended content decreased over time, leading to a convergence towards similar content and ultimately causing filter bubble effect. Ge \textit{et al.} \cite{under_EC} extended this analysis to e-commerce platforms, revealing a similar decline in content diversity. Michiels \textit{et al.} \cite{newsFB} further measured the filter bubble effect in news recommender systems. Nagulendra \textit{et al.} \cite{FBvis} simulated the emergence of filter bubble in social networks. Cinelli \textit{et al.} \cite{cinelli} analyzed social media platforms and found homogenization of users in Facebook. McKay \textit{et al.} \cite{McKay} explored user behavior on social media and corroborated the existence of filter bubbles. Aridor \textit{et al.} \cite{aridor} conducted simulations showing that users tend to consume increasingly narrower ranges of content over time. Li \textit{et al.} \cite{22nianli} identified that specific types of video content and user preference diversity cause filter bubbles in short-video platforms. Piao \textit{et al.} \cite{23piao} proposed that the origin of filter bubble lies in the diminishing diversity of information with mechanistic model. Michiels \textit{et al.} \cite{newsFB} used a regression model to verify the existence of filter bubbles through measures of topic and political variety. Chitra \textit{et al.} \cite{20Chitra} extended opinion dynamics models to investigate how content filtering algorithms contribute to societal polarization. Zhang \textit{et al.} \cite{23Zhang} simulated political polarization and found that users are rapidly exposed to more extreme content and become more radical in news recommendations.

These studies collectively affirm the presence of the filter bubble effect in recommender systems, which is traditionally understood as a problem of declining content diversity.

\subsubsection{Approaches to Mitigating Filter Bubble Effect in Recommender Systems}
A variety of approaches have been developed to address the filter bubble effect, with most focusing on enhancing diversity in recommendations. Symeonidis \textit{et al.} \cite{CouFB} increased novelty and diversity of recommendations using a Non-negative Matrix Factorization (NMF) approach. Sun \textit{et al.} \cite{PFB} proposed an adaptive diversity-regularized collaborative deep matrix factorization model to improve cross-domain recommendation diversity. Grossetti \textit{et al.} \cite{grossetti2019community} mitigated the filter bubble effect on the Twitter dataset based on community similarity. Gao \textit{et al.} \cite{gao22miFB} introduced a variational autoencoder-based recommender system that boosted diversity without compromising accuracy. Gao \textit{et al.} \cite{Gao} employed causal inference and reinforcement learning to counteract the overexposure effect in interactive recommender systems. Xu \textit{et al.} \cite{23Xu} tackled both the filter bubble and cold-start problems simultaneously, enhancing recommendation diversity and serendipity. Masrour \textit{et al.} \cite{20masrour} presented novel fairness-aware methods to address the filter bubble issue in network link prediction. Tommasel \textit{et al.} \cite{21Tommasel} developed an echo chamber-aware recommender system aiming at recommending novel and diverse users in social networks. Anand \textit{et al.} \cite{22mifb} analyzed existing deep recommender systems and enhanced them to improve recommendation diversity. Li \textit{et al.} \cite{li2023breaking} utilized reinforcement learning to add potential links between communities to alleviate the filter bubble effect caused by feedback loops.

\end{sloppypar}
\end{document}